\documentclass[reprint,amsmath,amssymb,aps,superscriptaddress,twocolumn,10pt]{revtex4-1}
\usepackage[colorlinks=true,allcolors=blue]{hyperref}
\usepackage{graphicx,color}
\usepackage{dcolumn}
\usepackage{subfigure}
\usepackage{bm}
\usepackage{amsmath,amsfonts,amssymb}
\usepackage{acronym}
\usepackage{enumitem}
\usepackage{array}
\usepackage{multirow}
\usepackage{CJK}
\allowdisplaybreaks
\newcommand{\be}{\begin{equation}}
\newcommand{\ee}{\end{equation}}
\newcommand{\bea}{\begin{eqnarray}}
\newcommand{\eea}{\end{eqnarray}}

\newcommand{\Msun}{{\rm M}_\odot}

\newcommand{\SOUTHCUT}{
School of Physics and Optoelectronics, South China University of Technology, Guangzhou 510641,
People's Republic of China}

\newacro{EMRI}{extreme mass-ratio inspirals}
\newacro{MBH}{massive black hole}
\newacro{BH}{black hole}
\newacro{GR}{general relativity}
\newacro{HKBH}{hairy Kerr black hole}
\newacro{KNBH}{Kerr-Newmann black hole}
\newacro{KBH}{Kerr black hole}
\newacro{NHT}{no-hair theorem}
\newacro{DWD}{double white dwarf}
\newacro{GW}{gravitational wave}
\newacro{AK}{analytic kludge}
\newacro{NK}{numerical kludge}
\newacro{AAK}{augmented analytic kludge}
\newacro{CO}{compact object}
\newacro{PE}{parameter estimation}
\newacro{SNR}{signal-to-noise ratio}
\newacro{PN}{post newtonion}
\newacro{FIM}{Fisher information matrix}
\newacro{LSO}{last stable orbit}
\newacro{ISCO}{innermost stable circular orbit}
\newacro{BBH}{Binary Black Hole}
\newacro{BNS}{Binary Neutron Star}
\newacro{NS}{Neutron Star}
\newacro{KN}{Kerr-Newmann}
\newcommand{\beq}{\begin{equation}}
\newcommand{\eeq}{\end{equation}}
\newcommand{\beqa}{\begin{eqnarray}}
\newcommand{\eeqa}{\end{eqnarray}}

\newcommand{\Op}{\hat{\mathcal{D}}_2}

\newcommand{\afn}{u}

\def\lsim{\mathrel{\rlap{\lower4pt\hbox{\hskip0.5pt$\sim$}}
    \raise1pt\hbox{$<$}}}         
\def\gsim{\mathrel{\rlap{\lower4pt\hbox{\hskip0.5pt$\sim$}}
    \raise1pt\hbox{$>$}}}         
\begin{document}
\begin{CJK*}{UTF8}{gbsn}
\title{Detecting the massive vector field with extreme mass-ratio inspirals}
\author{Tieguang Zi}
\affiliation{\SOUTHCUT}
\author{Chao Zhang}
\email{zhangchao666@sjtu.edu.cn, Corresponding author}
\affiliation{Institute of Fundamental Physics and Quantum Technology, Department of Physics, School of Physical Science and Technology, Ningbo University, Ningbo, Zhejiang 315211, China}

\begin{abstract}
The future space-borne gravitational wave detector, Laser Interferometer Space Antenna (LISA), has the potential of detecting the fundamental fields, such as the charge and mass of ultra-light scalar field.
In this paper we study the effect of lighter vector field on the gravitational waveforms from \ac{EMRI} system, consisting of a stellar-mass object and the \ac{MBH} in the Einstein-Proca theory of a massive vector field coupling to gravity.
Using the perturbation theory, we compute the energy fluxes including the contributions of the Proca field and the gravitational field, then obtain the adiabatic inspiraling orbits and corresponding waveforms.
Our results demonstrate that the vector charge and mass carried by the secondary body lead to detectable effects on EMRI waveform, and LISA has the potential to measure the mass of the Proca field with greater precision.
\end{abstract}
\maketitle
\end{CJK*}
\section{Introduction}
The fundamental theories not only predicate the existence of ultralight particles beyond standard model,
but also the possibility of the dark matter or dark energy.
The new particles described by the lighter vector fields, such as the dark photon,
generally exist in the modified theories \cite{Goodsell:2009xc,Essig:2013lka}.
These additional fundamental fields generally couple weakly to the baryonic matter in the standard model,
rendering their experimental detection extremely challenging.
However, there is a potential avenue for observing the effects of ultralight bosons in extreme, strongly gravitational environments using the mechanism of superradiant instability \cite{Brito:2015oca}.
The phenomenon could occur when the bosons wave frequencies $\omega_R$ meet the superradiant condition, that is $\omega_R\leq m \Omega_H$, where $\Omega_H$ is the horizon angular velocity and
$m$ is the azimuthal quantum number of the unstable mode.
When the condition of superradiant is satisfied, the massive fundamental fields, such as the scalar fields \cite{Dolan:2007mj,Arvanitaki:2016qwi,Arvanitaki:2014wva}
and the vector fields\cite{Pani:2012vp,Pani:2012bp,Witek:2012tr,East:2017ovw,East:2018glu,Dolan:2018dqv,Siemonsen:2019ebd},
can be condensed as the bosons clouds around the rotating \ac{BH}.
During the dissipation process of bosons clouds, the black hole-cloud system would produce the
quadrupolar GWs.
With the detection of such GW signal,
a number of works have started to search for the ultralight bosons with the
\ac{GW} detectors \cite{Arvanitaki:2014wva,Brito:2017zvb,Siemonsen:2019ebd,Jones:2023fzz,LIGOScientific:2024dzy,
Siemonsen:2019ebd,LIGOScientific:2024dzy,Michimura:2020vxn,Miller:2023kkd,Vermeulen:2021epa,Konoplya:2023fmh,Mitra:2023sny}.

Those fundamental fields predicted by the modified gravitational theories are regarded as the key components of some dark matter candidates,  where the particles of dark matter has a large range of masses in the different theories.
For example, the fuzzy dark matter particles has the mass as lower as $10^{-22}\rm eV$ \cite{Hui:2016ltb} and the mass of the primordial black holes can reach the sub-solar masses \cite{Miller:2020kmv}.
Additionally, the dark matter model predicates the dark photon with a mass as lower as $\sim 10^{-22}$eV, whose behaviors like the non-relativistic matter \cite{Chen:2016unw}.
The formation of such dark matter particles has many kinds of channels,
including the misalignment mechanism in a non-minimal coupling to gravity \cite{Nelson:2011sf}, quantum fluctuations during inflation \cite{Graham:2015rva} and appearing naturally in the string theories \cite{Goodsell:2009xc}.
Ones have carried out various kinds of  experimental detection schemes to probe the ultralight particles,
such as testing equivalence principle, where the constraints on the coupling effects of the ultralight particles to standard model
are implemented  by the E\"{o}t-Wash group \cite{Schlamminger:2007ht} and the Lunar Laser Ranging groups \cite{Williams:2004qba,Turyshev:2006gm}.
It may be advantageous that the GW detectors improve the limits on the ultralight fields.
The GW signal from the superradiance may allow LISA to constrain the vector field mass within
the range $10^{-16}-6\times 10^{-16}$eV \cite{Siemonsen:2022yyf}, and the EMRI signal including the
correction of the vector boson-cloud near the MBH can narrow the mass ranges of vector field,
reaching to $1.8\times 10^{-17}-4.47\times 10^{-16}$eV \cite{Fell:2023mtf}.

A promising target source of LISA is the \ac{EMRI} with a smaller mass-ratio $(\eta\in[10^{-7},10^{-4}])$, a stellar-mass body moving slowly close to MBH under the influence of the gravitational radiation backreaction, which can act as probing the fundamental theories and testing the nature of MBH \cite{Berry:2019wgg,Berti:2019xgr}.
For a typical EMRI system, the secondaries generally experience $\sim 10^{4}-10^{5}$ orbit cycles around the \ac{MBH}, so the inspiral orbits are inevitably involved with the relativistic effects in the strong gravitational field regime \cite{Berry:2019wgg,Berti:2019xgr}.
The unique feature of EMRI allows to conduct a series of testing \ac{GR} with the unprecedented precision \cite{Babak:2017tow,Zi:2021pdp,Cardenas-Avendano:2024mqp}, detecting the environmental effects near the MBH \cite{Cardoso:2022whc,Duque:2023cac,Speeney:2024mas,Rahman:2023sof,Cole:2022yzw,Zhang:2024ugv,Brito:2023pyl}, and ensures high precision estimation of the physical parameters of the EMRI system \cite{Berti:2019xgr,Babak:2017tow}, such as the mass and spin MBH \cite{Glampedakis:2005hs,Babak:2017tow,Fan:2020zhy}.
The accurate measurement of source parameters using the EMRI signal can allow to test accurately the
fundamental gravitational theories \cite{Berti:2015itd,Berti:2019xgr,Barack:2018yly,Cardenas-Avendano:2024mqp} and detect the rich environments neighbouring the MBH \cite{Bonga:2019ycj,Pan:2021ksp,Pan:2023wau,Cardoso:2022whc,Dai:2023cft,Figueiredo:2023gas,Zi:2023omh,Rahman:2023sof}.

Recently, EMRI as the probe of detecting fundamental fields beyond the \ac{GR} have been
broadly investigated in the Refs. \cite{Maselli:2020zgv,Maselli:2021men,Barsanti:2022ana,Barsanti:2022vvl,Zhang:2022rfr,
Zhang:2023vok,Fell:2023mtf,DellaRocca:2024pnm,Liang:2022gdk,Ghosh:2024arw},
where the MBH can be described by the Kerr metric according to the no-hair theorem and the stellar-mass \ac{BH}s can carry the scalar charges that are dependent on the masses \cite{Mignemi:1992nt,Kleihaus:2011tg,Antoniou:2017acq,Antoniou:2017hxj,Doneva:2017bvd,Maselli:2020zgv}.
Such assumption is based on the conclusion that the charge of BH is weaker when it becomes more massive.
It is due to that the charges are controlled by the curvature, and when the mass of BH is increasing the   curvature near the horizon tend to diminish \cite{Maselli:2020zgv}.
Therefore, the MBH would behave the much weaker deviation of Kerr hypothesis, and the stellar-mass BHs may occur the violation of the no-hair theorem.
In our paper, we focus on the EMRI system in the Einstein-Proca theory of a massive vector field coupled to gravity, where the theory and its generalized theories could exist the hairy BH solutions after setting the specific parameters
\cite{Chagoya:2016aar,Minamitsuji:2016ydr,Babichev:2017rti,Herdeiro:2016tmi,Heisenberg:2017xda,Kase:2018owh,Kase:2018voo}, boson star \cite{Minamitsuji:2017pdr} and vector star \cite{Kase:2017egk,Kase:2020yhw}.
Considering that the Einstein-Proca theories admit the same solutions in the GR,
and according to the method in the Refs. \cite{Maselli:2020zgv,Maselli:2020zgv,Maselli:2021men,Konoplya:2005hr,Konoplya:2006gq}, we can assume that the exterior spacetime of MBH are  approximatively described the Schwarzschild metric and the stellar-mass object carries the massive vector hair.

In our paper, we plan to constrain the Proca field mass using the waveform from EMRI system with
the vector-hairy stellar-mass object.
In Sec. \ref{method},  we present the massive vector field perturbations equation in the Schwarzschild spacetime and the perturbation source term of the point particle.
In Sec. \ref{waveform}, we introduce the progress of computing fluxes and analysis method of EMRI waveform.
Sec. \ref{result} shows our results including the Proca fluxes, dephasing and mismatch, the constraint on the mass of Proca field.
Finally, our conclusion and discussion are given in Sec. \ref{sumup}.
\section{Method}\label{method}
\subsection{Setup}
Firstly, we consider the action
\begin{equation}\label{action}
S(g,A,\psi) = S_0(g,A)+ \alpha S_c(g, \psi) + S_m(g,A,\psi)
\end{equation}
where $A$ is the Proca field, $g$ is the metric, $S_0(g,A)$ is the action relating to the spacetime background,
and $S_m(g,A,\psi)$ is the action describing the matter field $\psi$.
$\alpha S_c(g, \psi)$ denotes to the non-minimal couplings between the metric tensor $g$  and the Proca field $A$. $\alpha$ is a coupling constant, which has the dimensions $[\alpha]=(mass)^n$ with a positive integer $n$.
In this paper, we assume that the Proca field is minimally coupled to a supermassive black hole, the action can be given by
\begin{eqnarray}\label{Lagrangian}
S_0(g,A)&= &\int d^4x \sqrt{-g}
\left[\frac{c^4}{16\pi G} R - \mathcal{L}\right]
\end{eqnarray}
and
\begin{equation}
\mathcal{L} = \frac{1}{4}F_{\mu\nu}F^{\mu\nu}+ \frac{\mu_v^2}{2}A_{\mu}A^\mu,
\end{equation}
where $R$ is the Ricci scalar, $F_{\mu\nu}=\nabla_\mu A_{\nu}- \nabla_\nu A_{\mu}$ is the Proca field strength, $J^{\mu}$ is the current density for the massive vector and  $\mu_v$ is the mass of Proca field, which can be expressed with a dimensionless parameter $\mu$, that is $\mu_v=\mu M$.
The Proca mass can be written with the
dimension of $\rm eV$ \cite{Brito:2015oca,Fell:2023mtf}
\begin{equation}
\mu_v [{\rm eV}] \sim  \frac{\mu }{G\cdot 10^7 \Msun}  \hbar \cdot  c^3~
\rm eV,
\end{equation}
where the quantify $G$ is gravitational constant, $c$ is the speed of light and
$\hbar$ is the Planck constant.

For the EMRI system, the inspiraling secondary object with the mass $m_p$  can be regarded as
a moving point particle around the MBH with mass $M\gg m_p$.
Following the ``skletetonized" method adopted by the Refs. \cite{Ramazanoglu:2016kul,Maselli:2020zgv,Maselli:2021men,Barsanti:2022vvl},
the action of matter field $S_m(g,A,\psi)$ can be replaced with the action $S_p(g,A)$ of the point particle, that is
\be
S_m(g,A)\rightarrow S_p(g,A,\psi).
\ee
The action of a point particle is written as
\bea
S_p(g,A,\psi) &=& -\int m_p d\tau + \int d^4x \sqrt{-g} A_\mu J^\mu
\eea
where $J^\mu = q m_p \int d\tau \frac{dx^\mu}{d\tau} \frac{\delta^{(4)} (x-z(\tau))}{\sqrt{-g}}$ is the current density for vector field and $q$ is the charge of the Proca field.
From the action \eqref{action}, we can obtain the gravitational equation of motion
\begin{eqnarray}
G^{\mu\nu}=R^{\mu\nu} -\frac{1}{2} g^{\mu\nu} R&=& \mathcal{T}^{\mu\nu}_{\rm Proca} \nonumber  +  \mathcal{T}_p^{\mu\nu} + \alpha \mathcal{T}^{\mu\nu}_c \,,\\
\end{eqnarray}
where $ \mathcal{T}_p^{\mu\nu}=8\pi \int m_p \frac{\delta^{(4)} (x-y_p(\tau))}{\sqrt{-g}} \frac{dx_p^\mu}{d\tau} \frac{dx_p^\nu}{d\tau} d\tau $ is the energy-momentum tensor of the point particle,  $\mathcal{T}^{\rm Proca}_{\mu\nu}$ is the stress-energy tensor of the Proca field
\begin{equation}
\mathcal{T}_{\rm Proca}^{\mu\nu}  = 8\pi \left(-\frac{1}{4} F_{\rho\sigma} F^{\rho\sigma} g^{\mu\nu} + F_{~\rho}^{\mu} F^{\rho\nu}  - \frac{1}{2} \mu^2 g^{\mu\nu}A_\rho A^\rho +\mu^2 A^\mu A^\nu \right),
\end{equation}
and
$\mathcal{T}^{c}_{\mu\nu}=- \frac{16\pi}{\sqrt{-g}} \frac{\delta S_c}{\delta g^{\mu\nu}}$.
The Proca field equation can be written as
\be\label{procaequation}
\nabla_\rho F^{\rho\mu}-\mu^2 A^\mu + \frac{8\pi\alpha}{\sqrt{-g}} \frac{\delta S_c}{\delta A^\mu} = 8\pi J^\mu .
\ee

Given that the coupling constant $\alpha$
and the mass-ratio $\eta$ of EMRI has as following relationship, $\alpha/M^n = (\alpha/m_p^n)\eta^n$,
and the current observations do not support  the deviation from Kerr black holes  \cite{Nair:2019iur},
so we can assume that $\alpha/m_p^n$ is far less than one \cite{Barsanti:2022vvl}.
Thus the quantities $\alpha\frac{\delta S_c}{\delta A_\mu} $  and $\alpha \mathcal{T}_{\mu\nu}^c$ are both
ignorable \cite{Maselli:2020zgv}.
When the coupling constant $\alpha \rightarrow 0$, the no-hair theorems still hold in the
theories with the Proca fields \cite{Garcia-Saenz:2021uyv}.
If considering only the leading order of the vector perturbations in the
mass-ratio, the fields in the exterior spacetime of the MBH tend to a constant $A_0$,
the $\mathcal{T}_{\mu\nu}^{\rm Proca}$ can be reduced as a quadratic term around $A_0$, which is also overlooked as the case of the scalar field perturbations in the Refs.\cite{Maselli:2020zgv,Maselli:2021men,Barsanti:2022vvl}.
In conclusion, in the framework of approximations,
we can assume that \ac{MBH} is approximatively described by the Schwarzchild metric,
and the point particle, carrying the vector charge and mass,
acts as the Proca field with the Eq. \eqref{procaequation}.
Finally, the gravitational field equation can be simplified as
the GR case, the right-hand side of the Proca field equation acts as the source terms.

The spacetime metric of the Schwarzchild MBH can be written as
\begin{eqnarray}
ds^2&=&g_{\mu\nu}dx^\mu dx^\nu \nonumber\\
&=&-f(r) dt^2 + \frac{dr^2}{f(r)} +r^2 \left( d\theta^2 + \sin^2 \theta d\phi^2 \right)
\end{eqnarray}
with $f(r) = 1 - 2M/r$.
Due to the presence of spherical symmetry, the Proca equation can be decomposed as four second-order partial differential equations using vector harmonics \cite{Rosa:2011my}. The four-vector harmonics $Z^{(i)lm}_{\mu}$  can given by following
\begin{eqnarray} \label{vector-harmonics}
Z_{\mu}^{(1)lm} &=& \left[ 1, 0, 0, 0 \right] Y^{lm}, \\
Z_{\mu}^{(2)lm} &=& \left[ 0, f^{-1}, 0, 0 \right] Y^{lm}, \\
Z_{\mu}^{(3)lm} &=& \frac{r}{\sqrt{l(l+1)}} \left[ 0, 0, \partial_\theta, \partial_\phi  \right] Y^{lm}, \\
Z_{\mu}^{(4)lm} &=&  \frac{r}{\sqrt{l(l+1)}} \left[ 0, 0, \frac{1}{\sin\theta} \partial_\phi, - \sin\theta~\partial_\theta \right] Y^{lm}~,
\end{eqnarray}
where $Y^{lm} \equiv Y^{lm}(\theta, \phi)$ denotes to the ordinary spherical harmonics,
which satisfy the following orthogonality condition
\beq
\int \left(Z_{\mu}^{(i)lm}\right)^\ast \eta^{\mu \nu} Z_\nu^{(i')l'm'} d\Omega = \delta_{ii'} \delta_{ll'} \delta_{mm'}~,
\eeq
where $\eta^{\mu\nu}=\text{diag}[1, f^2, 1/r^2, 1/(r^2\sin^2\theta)]$ and $d\Omega=\sin\theta d\theta d\phi$.
Using the vector spherical harmonics, the vector potential and current density can be decomposed into the Fourier harmonic components
\begin{eqnarray}\label{ansatz}
A_{\mu}(t,r,\theta,\phi) &=&\frac{1}{r}\int_{-\infty}^{+\infty}d\omega~ \sum_{i=1}^{4} \sum_{lm} c_i \, u^{lm}_{(i)}(\omega,r) Z_\mu^{(i)lm}(\theta, \phi)
\nonumber \\ && e^{-i\omega t}~,\\
J_{\mu}(t,r,\theta,\phi) &= &\frac{1}{r}\int_{-\infty}^{+\infty}d\omega~ \sum_{i=1}^{4} \sum_{lm} c_i \, s^{lm}_{(i)}(\omega,r) Z_\mu^{(i)lm}(\theta, \phi)
\nonumber \\ && e^{-i\omega t}~,
\end{eqnarray}
where $c_1 = c_2 = 1$, $c_3 = c_4 = [l(l+1)]^{-1/2}$.
Then the Proca field equation (\ref{procaequation}) can be transformed into a set of four second-order partial differential equations using the ansatz \eqref{ansatz}
\begin{eqnarray}
\Op \afn_{(1)} &+& \left[ \frac{2}{r^2} \left(  \dot{\afn}_{(2)} - \afn^\prime_{(1)} \right) \right] = 4\pi f s_{(1)} ,\label{eq-alp1} \\
\Op \afn_{(2)} &+& \frac{2}{r^2}\left[ \left(  \dot{\afn}_{(1)} - \afn^\prime_{(2)} \right) - f^2\left( \afn_{(2)} - \afn_{(3)} \right) \right]
\nonumber \\ &=&4\pi f s_{(2)},  \label{eq-alp2}\\
\Op \afn_{(3)} &+& \left[ \frac{2 f l (l+1)}{r^2} \afn_{(2)} \right] = 4\pi f s_{(3)}, \label{eq-alp3}  \\
\Op \afn_{(4)} &=& 4\pi f s_{(4)}~,   \label{odd-parity}
\end{eqnarray}
where $\dot\afn \equiv \tfrac{\partial \afn}{\partial t}$, $\afn^\prime \equiv \tfrac{\partial \afn}{\partial r^\ast}$, the tortoise coordinate $r_\ast = r+2\ln\left(r/2M-1\right)$, and the differential operator $\Op$ is
\beq
\Op \equiv \frac{\partial^2}{\partial r_\ast^2}+ \omega^2 - f \left[ \frac{l(l+1)}{r^2} + \mu^2 \right]~.
\eeq

The fourth equation (\ref{odd-parity}), describing the odd-parity sector, is completely decoupled from the first three equations  (\ref{eq-alp1})--(\ref{eq-alp3} that describe the even-parity sector. The even-parity equations have the following relationship in term of the Lorenz condition \cite{Rosa:2011my},
\beq
-\dot{\afn}_{(1)} + \afn^\prime_{(2)} + \frac{f}{r} \left(\afn_{(2)} - \afn_{(3)}  \right)  =  0~.  \label{Lorenz-eq}
\eeq
Using the condition \eqref{Lorenz-eq}, the even-parity system can be reduced to a pair of coupled differential equations, then the Eq.~(\ref{eq-alp2}) is recast as
\begin{eqnarray}
\Op \afn_{(2)} - \frac{2 f}{r^2} \left(1 - \frac{3}{r} \right) \left( \afn_{(2)} - \afn_{(3)} \right) = 4\pi f s_{(2)}  \label{eq-alp2-alt} .
\end{eqnarray}
Thus, the four equations (\ref{eq-alp1})-\ref{odd-parity}) finally are reduced as one decoupled wave equation for the odd-parity part and two coupled wave equations for the even-parity mode.

\subsection{Source term of point particle}
In this paper, the topic of our research is the dynamic of EMRI, so we focus on the perturbation term of
point particle around the MBH.
Assuming the trajectories of the inspiraling point particle are the circular geodesic orbits on the equatorial plane, that is $(r_p=r_0,\, \theta_p=\pi/2,\, \phi_p=\Omega_p \, t)$, the current density for the Proca field can be written as
\begin{equation}
J^\mu=q\frac{u^\mu}{u^t r_p^2 \sin\theta_p}\delta(r-r_p)\delta(\theta-\theta_p)\delta(\phi-\phi_p),
\end{equation}
where $q$ is the vector charge  and $u^\mu$ is the velocity of the particle
\begin{equation}
u^{\mu}=\left(E_p/f_p,0,0,L_p/r_p^2\right).
\end{equation}
Here the particle's orbital energy $E_p$, angular momentum  $L_p$ and angular velocity $\Omega_p$ in the Schwazchild spacetime are \cite{Poisson:1993vp}
\begin{equation}
\begin{split}
E_p&=f_p(1-3/r_p)^{-1/2},  \\
L_p&=r_p^{1/2}(1-3/r_p)^{-1/2}, \\
\Omega_p&=\sqrt{M}r_p^{-3/2},
\end{split}
\end{equation}
with $f_p=1-2M/r_p$.
The source terms for perturbed equations Eqs. \eqref{eq-alp2}-\eqref{odd-parity} are given by
\begin{equation}
\begin{split}
s_{(2)}&=0,\\
s_{(3)}&=-q\frac{\partial_\phi Y^{lm}(\pi/2,0)}{r_p^{3/2}}\delta(r-r_p)\delta(\omega-\omega_m),\\
s_{(4)}&=-q \frac{\partial_\theta Y^{lm}(\pi/2,0)}{r_p^{3/2}}\delta(r-r_p)\delta(\omega-\omega_m),
\end{split}
\end{equation}
with the orbital frequency $\omega_m=m\Omega_p$.

\section{Fluxes and Waveform}\label{waveform}
In order to compute the massive vector field emission of a charged particle around the Schwarzchild BH, we shall adopt the Green's function approach to compute the fluxes for the Proca field, the first procedure is to obtain homogeneous solutions of the equations (\ref{eq-alp3}) and (\ref{eq-alp2-alt}) for the even-parity mode as well as the equation (\ref{odd-parity}) for the odd-parity mode, then compute the fluxes using the homogeneous solutions and the source terms.
\subsection{Odd sector}
We first start to consider solving the odd-part equation (\ref{odd-parity}) with the
numerical method, which has the following boundary conditions
\begin{eqnarray}
u_{(4)}^{H}&=&\sum_{n=0}^{n_h}e^{-i \omega r_\ast(r)}a_n(r-r_h)^n,\\
u_{(4)}^{\infty}&=&\sum_{n=0}^{n_\infty}e^{i \sqrt{\omega^2-\mu^2} r_\ast(r)}r^{\frac{i\mu^2}{\sqrt{\omega^2-\mu^2}}}\frac{b_n}{r^n},
\end{eqnarray}
then obtain the ingoing near the horizon and the outgoing solutions at the spatial infinity, respectively.
One can get the coefficients $(a_n,b_n)$ by solving the homogeneous Eq. \eqref{odd-parity} at each order in $(r-r_h)$ and $1/r$, where the coefficients of the leading terms $a_0=b_0=1$, and the highest orders of expansion expression are set to $n_h=n_\infty=n_{\rm max}=8$.
These boundary conditions with the higher-order expansion expression can ensure accurate numerical integration.

Then the general solution  for the odd sector can be given as follows
\begin{equation}
    u_{(4)}=u_{(4)}^{H}\int_{r}^{\infty}d r_\ast~ \frac{4\pi f s_{(4)}}{W}u_{(4)}^{\infty}+u_{(4)}^{\infty}\int_{-\infty}^{r}d r_\ast~ \frac{4\pi f s_{(4)}}{W}u_{(4)}^{H} \, ,
\end{equation}
with the wronskian
\begin{equation}
W=\frac{du_{(4)}^{\infty}}{d r_\ast}u_{(4)}^{H}- \frac{du_{(4)}^{H}}{d r_\ast}u_{(4)}^{\infty}.
\end{equation}
We can compute the general solutions in terms of the homogeneous solutions $u_{(4)}^{\infty,H}$ at infinity and near the horizon
\begin{equation}
\begin{split}
\lim_{r_\ast\to -\infty}u_{(4)}&=e^{-i \omega r_\ast(r)}\int_{-\infty}^{\infty}d r_\ast~\frac{4\pi f s_{(4)}}{W} u_{(4)}^{\infty}\\
&=\mathcal{A}^{4-}_{lm}e^{-i \omega r_\ast(r)}, \\
\lim_{r_\ast\to +\infty}u_{(4)}&= e^{i \sqrt{\omega^2-\mu^2} r_\ast(r)}r^{\frac{i\mu^2}{\sqrt{\omega^2-\mu^2}}}\int_{-\infty}^{\infty}d r_\ast~\frac{4\pi f s_{(4)}}{W} u_{(4)}^{H} \\
&=\mathcal{A}^{4+}_{lm}e^{i \sqrt{\omega^2-\mu^2} r_\ast(r)}r^{\frac{i\mu^2}{\sqrt{\omega^2-\mu^2}}}.
\end{split}
\end{equation}

\subsection{Even sector}
In this subsection, we consider the calculation of Proca fluxes for the even-parity modes
by computing the homogeneous solutions of the Eqs. (\ref{eq-alp3}) and (\ref{eq-alp2-alt})
and the source term.
The ordinary differential equations (ODE) can be recasted into the equation matrix just as Ref. \cite{Zhu:2018tzi}
\begin{align}
\label{eq:master}
\left( \frac{d^2}{dr_*^2} + \omega^2 + \left[\begin{array}{cc} \alpha_{lm} & \beta_{lm} \\ \gamma_{lm} & \sigma_{lm} \end{array} \right] \right) \left[\begin{array}{c} u_{(2)} \\ u_{(3)} \end{array} \right] &= \left[\begin{array}{c} S_{lm} \\ Z_{lm} \end{array} \right],
\end{align}
where the coefficients $\alpha_{lm}$, $\beta_{lm}$, $\gamma_{lm}$, $\sigma_{lm}$ are
\begin{equation}
\begin{split}
    \alpha_{lm}&=-f \left[ \frac{l(l+1)}{r^2} + \mu^2 \right]-\frac{2 f}{r^2} \left(1 - \frac{3}{r} \right),\\
\beta_{lm}&=\frac{2 f}{r^2} \left(1 - \frac{3}{r} \right),\\
\gamma_{lm}&=\frac{2 f l (l+1)}{r^2},\\
\sigma_{lm}&=-f \left[ \frac{l(l+1)}{r^2} + \mu^2 \right].
\end{split}
\end{equation}
It should be noted that these ODE coefficients satisfy the asymptotic behaviors
near the horizon $r=r_h$ and at the infinity  $r=\infty$
\begin{eqnarray}\label{rhor:even:asy}
\underset{r_* \rightarrow -\infty\;\;\;\;\;\;}{\text{lim}\;\, \alpha_{lm}}& =& \underset{r_* \rightarrow -\infty\;\;\;\;\;\;}{\text{lim}\;\, \beta_{lm}} = 0 ,  \nonumber \\
\underset{r_* \rightarrow -\infty\;\;\;\;\;\;}{\text{lim}\;\, \gamma_{lm}}& =&\underset{r_* \rightarrow -\infty\;\;\;\;\;\;}{\text{lim}\;\, \sigma_{lm}} = \; 0,
\end{eqnarray}
and
\begin{eqnarray} \label{rinf:even:asy}
\underset{r_* \rightarrow +\infty\;\;\;\;\;\;}{\text{lim}\;\, \beta_{lm}} &=&\underset{r_* \rightarrow +\infty\;\;\;\;\;\;}{\text{lim}\;\, \gamma_{lm}}  = \; 0
\nonumber  \\ \underset{r_* \rightarrow +\infty\;\;\;\;\;\;}{\text{lim}\;\, \alpha_{lm}}& =&\underset{r_* \rightarrow +\infty\;\;\;\;\;\;}{\text{lim}\;\, \sigma_{lm}} = \; -\mu^2.
\end{eqnarray}
Using the above-mentioned conditions \eqref{rhor:even:asy} and \eqref{rinf:even:asy},
one can obtain four independent homogeneous solutions by solving the master equations \eqref{eq:master}.

In the following section, every homogeneous solution with a superscript
denotes the asymptotic behaviors at the boundary.
Concretely, for the infinity $r_*=+\infty$, the two outgoing homogeneous solutions when $r\ll |\omega_m|^{-1}$ can be written as
\begin{eqnarray}
\label{eq:out}
\left[ \begin{array}{c} u_{(2)}^{0+} \\ u_{(3)}^{0+} \end{array} \right] \simeq e^{i \sqrt{\omega^2-\mu^2} r_\ast(r)}r^{\frac{i\mu^2}{\sqrt{\omega^2-\mu^2}}} \left[ \begin{array}{c} 1 \\ 0 \end{array} \right] ,\nonumber \\
 \;\;\;\;\; \left[ \begin{array}{c} u_{(2)}^{1+} \\ u_{(3)}^{1+} \end{array} \right] \simeq e^{i \sqrt{\omega^2-\mu^2} r_\ast(r)}r^{\frac{i\mu^2}{\sqrt{\omega^2-\mu^2}}} \left[ \begin{array}{c} 0 \\ 1 \end{array} \right] ,
\end{eqnarray}
and the two ingoing homogeneous solutions for the horizon $r_*=-\infty$ when $r-r_+ \ll M$ can
be given by
\begin{eqnarray}
\label{eq:down}
\left[ \begin{array}{c} u_{(2)}^{0-} \\ u_{(3)}^{0-} \end{array} \right] \simeq e^{-i\omega r_*} \left[ \begin{array}{c} 1 \\ 0 \end{array} \right] , \;\;\;\;\; \left[ \begin{array}{c} u_{(2)}^{1-} \\ u_{(3)}^{1-} \end{array} \right] \simeq e^{-i\omega r_*} \left[ \begin{array}{c} 0 \\ 1 \end{array} \right] .
\end{eqnarray}
Since the set of outgoing and ingoing homogeneous solutions is not unique,
using the linear combinations of Eqs. \eqref{eq:out} and \eqref{eq:down},
we can recast a new basis of homogeneous solutions.
It indeed is convenient to compute the inhomogeneous solutions with such a set basis.
Note that the homogeneous solutions are expanded as the power series near the boundary (either $r\rightarrow \infty$ or $r \rightarrow r_h$), and one can compute the initial values at the boundaries for Eq. \eqref{eq:master}, then obtain the global homogeneous solutions using the numerical method. The full details of the boundary expansions can be seen in \cite{Zhu:2018tzi}.

For the source term in the Eq. \eqref{eq:master}, one can express the following form using the Dirac delta functions
\begin{eqnarray}
\label{eq:source}
\left[\begin{array}{c} S_{lm}(r) \\ Z_{lm}(r) \end{array} \right] &=& \left[\begin{array}{c} B_{lm} \\ D_{lm} \end{array} \right]\delta(r-r_0) \nonumber \\ &+& \left[\begin{array}{c} F_{lm} \\ H_{lm} \end{array} \right] \frac{d}{dr}\delta(r-r_0) ,
\end{eqnarray}
where $B_{lm}$, $D_{lm}$, $F_{lm}$, and $H_{lm}$ can be given by the orbital elements of the point particle.
In terms of the source terms and homogeneous solutions, the inhomogeneous solution can be expressed as a piecewise function as following \cite{Zhu:2018tzi}
\begin{align}
\label{eq:inhomo}
&\left[\begin{array}{c} u_{(2)} \\ u_{(3)} \end{array} \right] = \left( \mathcal{A}^{2+}_{lm} \left[ \begin{array}{c} u_{(2)}^{0+} \\ u_{(3)}^{0+} \end{array} \right] + \mathcal{A}^{3+}_{lm} \left[ \begin{array}{c} u_{(2)}^{1+} \\ u_{(3)}^{1+} \end{array} \right]  \right) \Theta(r-r_0) \notag
\\&\qquad\;\;\; +\left( \mathcal{A}^{2-}_{lm} \left[ \begin{array}{c} u_{(2)}^{0-} \\ u_{(3)}^{0-} \end{array} \right] + \mathcal{A}^{3-}_{lm} \left[ \begin{array}{c} u_{(2)}^{1-} \\ u_{(3)}^{1-} \end{array} \right]  \right) \Theta(r_0-r) ,
\end{align}
where $\Theta$ is the Heaviside step function.  The normalization coefficients are satisfied by the following linear system (involving the Wronskian matrix)
\begin{align}
&\left[\begin{array}{cccc} h_{lm}^{0+} & h_{lm}^{1+} & h_{lm}^{0-} & h_{lm}^{1-}
\\ a_{lm}^{0+} & a_{lm}^{1+} & a_{lm}^{0-} & a_{lm}^{1-}
\\ \partial_{r_*} h_{lm}^{0+} & \partial_{r_*} h_{lm}^{1+} & \partial_{r_*} h_{lm}^{0-} & \partial_{r_*} h_{lm}^{1-}
\\ \partial_{r_*} a_{lm}^{0+} & \partial_{r_*} a_{lm}^{1+} & \partial_{r_*} a_{lm}^{0-} & \partial_{r_*} a_{lm}^{1-}  \end{array} \right]_{r_0} \left[\begin{array}{c} C_{lm}^{0+} \\ C_{lm}^{1+} \\ -C_{lm}^{0-} \\ -C_{lm}^{1-} \end{array} \right] \notag
\\&\qquad\qquad\;\;\; = \frac{1}{r_0^3 f_p^2} \left[\begin{array}{c} r_0^3 F_{lm} \\ r_0^3 H_{lm} \\ r_0^3 f_p B_{lm}+2Mr_0 F_{lm} \\ r_0^3 f_p D_{lm}+2Mr_0 H_{lm} \end{array} \right] ,
\label{eq:wron}
\end{align}
where all functions related to $r$ can be computed at $r=r_0$.

\subsection{Fluxes and Evolution}
Because we have assumed that the inspiraling orbits of secondaries are along the circular geodesics on the equatorial plane, under the condition of adiabatic approximation, the evolution of the orbital radius is approximately updated by the EMRI fluxes.
In this section, we focus on the energy fluxes from the EMRI system, which mainly
consist of the contributions from the Proca field and gravitational field.

First, we consider the Proca flux, which can be derived from Isaacson's stress-energy tensor \cite{Isaacson:1968zza}
\begin{equation}
4\pi T_{\mu\nu}= g^{\alpha\beta}F_{\mu \alpha}F_{\nu \beta}+\mu^2 A_{\mu}A_{\nu}-g_{\mu\nu}\mathcal{L}.
\end{equation}
Due to the static and axially symmetric of the Schwarzschild black hole, there are two Killing vectors
$\xi^\alpha_{(t)}$ and $\xi^\alpha_{(\phi)}$,
so one can compute the energy flux through a surface $\Sigma$ at constant $r$.
\be
d E = -\int_\Sigma T^\mu_{~\nu}~ _{(t)}\xi^\nu d\Sigma_\mu
\ee
where $d\Sigma_\mu$ is an outward oriented surface element on the surface $\Sigma$.
In Schwarzschild coordinates, the expression of the energy flux can be simplified as following \cite{Martel:2003jj}
\begin{equation} \label{dEdt:Formular}
\frac{dE^{ \mathcal{P}}}{dt}=-\epsilon r^2 f \int d\Omega~T_{tr},
\end{equation}
where the parameter $\epsilon$ is $1$ corresponding to the flux at the infinity and $-1$ corresponding to the flux near the horizon.
After some simplification using the formula \eqref{dEdt:Formular}, the energy flux formulas at infinity for the even and odd sectors are given by
\begin{equation}\label{eq:Edot:P:I:odd}
\left\langle \frac{d E ^{ \mathcal{P}}}{dt}\right\rangle_{\rm odd}^{\infty}=\sum_{l=1}^\infty \sum_{m=1}^l \frac{\omega_m\sqrt{\omega_m^2-\mu^2}|\mathcal{A}^{4+}_{lm}|^2}{2\pi l(l+1)}\,,
\end{equation}
\begin{eqnarray}\label{eq:Edot:P:I:even}
\left\langle \frac{d E ^{ \mathcal{P}}  }{dt}\right\rangle_{\rm even}^{\infty}&=&\sum_{l=1}^\infty \sum_{m=1}^l \frac{\omega_m\sqrt{\omega_m^2-\mu^2}|\mathcal{A}^{3+}_{lm}|^2}{2\pi l(l+1)}\\ \nonumber &+&\frac{\mu^2}{2\pi}\frac{\sqrt{\omega_m^2-\mu^2}}{\omega_m}|\mathcal{A}^{2+}_{lm}|^2\,.
\end{eqnarray}
The energy flux formulas near the horizon can be given by
\begin{equation}\label{eq:Edot:P:H:odd}
\left\langle \frac{d E ^{ \mathcal{P}}}{dt}\right\rangle_{\rm odd}^{H}=\sum_{l=1}^\infty \sum_{m=1}^l \frac{\omega_m^2|\mathcal{A}^{4-}_{lm}|^2}{2\pi l(l+1)}\,,
\end{equation}
\begin{eqnarray}\label{eq:Edot:P:H:even}
\left\langle \frac{d E^{ \mathcal{P}}}{dt}\right\rangle_{\rm even}^{H} &=&
\sum_{l=1}^\infty \sum_{m=1}^l \frac{\omega_m^2|\mathcal{A}^{3-}_{lm}|^2}{2\pi l(l+1)}
\\ \nonumber &+&\frac{l(l+1)+4\mu^2}{8\pi}|\mathcal{A}^{2-}_{lm}|^2
\\ \nonumber &-&\frac{i\omega_m}{4\pi}\left(\mathcal{A}^{2-}_{lm} \mathcal{A}^{3-\ast}_{lm}-\mathcal{A}^{3-}_{lm} \mathcal{A}^{2-\ast}_{lm}\right)\,.
\end{eqnarray}
The full details of the derivation for the flux formulas have been shown in the Appendix. \ref{appendix1}.
The angle brackets $<>$ mean the time average, and the symbols will be omitted to write concisely in the following section.
Based on the fluxes in the Eqs. \eqref{eq:Edot:P:I:odd}-\eqref{eq:Edot:P:H:even},
the total Proca flux can be given by
\begin{eqnarray}
\dot{E}^{\mathcal{P}} = \dot{E}^{\mathcal{P},H}_{\rm odd} + \dot{E}^{\mathcal{P},H}_{\rm even}
+ \dot{E}^{\mathcal{P},\infty}_{\rm odd} + \dot{E}^{\mathcal{P},\infty}_{\rm even},
\end{eqnarray}
where the overdot indicates the time derivative.

In the following section, we will consider the gravitational fluxes in the Schwarzchild spacetime, which can be computed by the Regge-Wheeler equation
and have been broadly studied for the odd sector and the even sector in the Refs. \cite{Regge:1957td,Zerilli:1970se,Martel:2003jj}.
The package of computing fluxes with the Regge-Wheeler equation is also available at the website of  \textit{Black Hole Perturbation Toolkit} \cite{BHPToolkit},
which can generate the gravitational flux $\dot{E}^\mathcal{G}$.
Thus, the total energy flux for the EMRI system can be regarded as the sum of the gravitational flux and the Proca flux
\begin{equation}
\dot{E} = \dot{E}^\mathcal{G} +\dot{E}^\mathcal{P},
\end{equation}
which will be used to evolve the point particle on the circular geodesic orbits.

In the framework of adiabatic approximation, the orbital loss energy results from the
gravitational and Proca emissions. Thus the evolution of the inspiraling point particle's orbital radius and frequency are derived by
\begin{equation} \label{equation:evolution:insp}
\frac{dr}{dt} = - \dot{E}\left(\frac{dE_{p}}{dr}\right)^{-1} , \quad \frac{d\Phi}{dt} = \Omega_p,
\end{equation}
where $\Phi$ is the orbital phase of the point particle.
One can obtain the orbital radius and frequency of the point particle by numerically integrating the Eqs. \eqref{equation:evolution:insp}, then compute the adiabatic EMRI waveform.
To assess the effect of the Proca field on the EMRI orbital dynamic, a rough rule can be considered by defining the dephasing, which can be given by
\be\label{dephasing:formula}
\delta \Phi(t) =  \Phi(t)^{\rm q,\mu=0} - \Phi(t)^{\rm q,\mu\neq0},
\ee
where $\Phi(t)^{\rm q,\mu=0}$ and $\Phi(t)^{\rm q,\mu\neq0}$ correspond to the orbital phase for the
inspiral evolution for the massless vector field case and the massive vector field case.

\subsection{Waveform data analysis}
In this subsection, we present the recipe of computing the waveform with the quadrupole formula, then
introduce the dephasing and mismatch to assess the effect produced by the Proca field on the EMRI waveform.

The expressions of GW waveform from the EMRI system in the quadrupole approximation \cite{Barack:2003fp,Huerta:2011kt,Jiang:2021htl}
can be given by
\begin{equation}
h_+=\frac{4\Omega_p^2 r^2}{d_L}\frac{1+\cos^2\iota}{2}\cos\left[2\varphi(t)\right],
\end{equation}
\begin{equation}
h_\times=\frac{4\Omega_p^2 r^2}{d_L}\cos\iota\sin\left[2\varphi(t)\right],
\end{equation}
where $d_L$ is the distance from the source to the detector, $r$ is the orbital radius of point particle, and $\iota$ describes the angle between the line-of-sight and the rotational axis of the orbits.
$\varphi(t)$ is the modulated GW phase due to the orbital motion of LISA \cite{Babak:2006uv}, which
can be written as
\begin{equation}\label{doppler}
\varphi(t)\to \varphi(t)+\varphi'(t)R_{\text{AU}}\sin\theta_s\cos\left(\frac{2\pi t}{T} -\phi_s-\phi_{\alpha}\right),
\end{equation}
where $\phi_{\alpha}$ is the ecliptic longitude of the detector $\alpha$ at $t=0$,
the rotational period $T$ is one year and  $R_{\text{AU}}$ is the astronomical unit.

Under the low frequency approximate, the EMRI waveform responded by the detector can be written as
\begin{equation}\label{signal}
h(t)= \frac{\sqrt{3}}{2}[h_{+}(t) F^{+}(t)+h_{\times}(t) F^{\times}(t)],
\end{equation}
where the antenna pattern functions $F^{+,\times}(t)$ and $\iota$ can be expressed in terms of the source orientation $(\theta_s,\phi_s)$ and the direction of the angular momentum $(\theta_1,\phi_1)$.
The signal-to-noise ratio (SNR) of the GW signals is defined by
\begin{equation}
\rho=\sqrt{\left\langle h|h \right\rangle},
\end{equation}
and the noise-weighted inner product between two templates $h_1$ and $h_2$ is
\begin{equation}\label{product}
\left\langle h_{1} \mid h_{2}\right\rangle=4 \Re \int_{f_{\min }}^{f_{\max }} \frac{\tilde{h}_{1}(f) \tilde{h}_{2}^{*}(f)}{S_{n}(f)} df,
\end{equation}
where
\begin{equation}
f_{\text{max}}=f_{\text{ISCO}},~~~~~~f_{\text{min}}=0.1 \rm mHz,
\end{equation}
$f_{\text{ISCO}}$ is the orbital frequency of \ac{ISCO} in the Schwarzchild spacetime, $\tilde{h}(f)$ is the Fourier transform of the time-domain signal $h(t)$,
its complex conjugate is $\tilde{h}^{*}(f)$,
and $S_n(f)$ is the noise spectral density of the space-based GW detectors, such as LISA \cite{LISA:2017pwj} and TianQin \cite{TianQin:2020hid,Gong:2021gvw}.

To assess the detection potential of the Proca field mass with the EMRI observation by LISA,
we carry out the parameter estimation with the FIM method.
For the EMRI waveform including the modification of the Proca field, the GW signal is mainly determined by the ten parameters
\begin{equation}
\xi=( M,  m_p, \mu_v, r_0, q,\theta_s, \phi_s, \theta_l, \phi_l, d_L),
\end{equation}
where $r_0$ is the initial radius of the orbit for the secondary compact object.
The angles $(\theta_s, \phi_s)$ and  $(\theta_l, \phi_l)$ denote to the orientation of EMRI system
and the direction of the orbital angular momentum in the barycentric frame \cite{Maselli:2021men}.

In the large SNR limit,
the covariances of source parameters $\xi$  are given by the inverse of the \ac{FIM}
\begin{equation}
{\bf \Gamma}_{i j}=\left\langle\left.\frac{\partial h}{\partial \xi_{i}}\right| \frac{\partial h}{\partial \xi_{j}}\right\rangle_{\xi=\hat{\xi}}.
\end{equation}
The statistical errors on $\xi$ and the correlation coefficients between the source parameters are provided by the diagonal and non-diagonal parts of ${\bf \Sigma}={\bf \Gamma}^{-1}$
\begin{equation}
\sigma_{i}=\Sigma_{i i}^{1 / 2}.
\end{equation}
Because of the triangle configuration of the space-based GW detector regarded as a network of two L-shaped detectors, with the second interferometer rotated of $60^\circ$ with respect to the first one, the total \ac{SNR} can be written as the sum of SNRs of two L-shaped detectors \cite{Cutler:1994pb}
\begin{equation}
\rho=\sqrt{\rho_1^2+\rho_2^2}=\sqrt{\left\langle s_1|s_1 \right\rangle+\left\langle s_2|s_2 \right\rangle},
\end{equation}
where $s_1$ and $s_2$ denote the signals detected by two L-shaped detectors.
The total covariance matrix of the source parameters is obtained by inverting the sum of the fisher matrices $\sigma_{\xi_i}^2=(\Gamma_1+\Gamma_2)^{-1}_{ii}$.

To evaluate the effect of the Proca field mass on the EMRI waveform,
one can  compute the mismatch between two signals, which is defined by the overlap
$\mathcal{O}$
\begin{equation}\label{eq:def_F}
\mathcal{M}\equiv1-\mathcal{O} = 1- \frac{<h_a|h_b>}{\sqrt{<h_a|h_a><h_b|h_b>}} .
\end{equation}
An empirical formula for distinguishing two kinds of waveforms is proposed \cite{Flanagan:1997kp,Lindblom:2008cm}, in particular the
detector can recognize two waveforms when their mismatch satisfies the conditions $\mathcal{M} \geq 1/(2\rho^2)$.
The threshold value of the \ac{SNR} of EMRI signals detected by the LISA or TianQin is conservatively chosen as $\rho=20$ \cite{Babak:2017tow,Fan:2020zhy}, so the threshold of the mismatch discerned by LISA  is  $\mathcal{M}_{\rm min}=0.00125$.

\section{Result}\label{result}
In this section, our results are presented in the following three parts: the flux of the Proca field in the Sec. \ref{fluxes}, the dephasing and mismatch for the EMRI waveforms including the modification of massless and massive vector fields in the Sec. \ref{dephasing}, and the constraint on the mass of Proca field using the FIM method in the Sec. \ref{fim}.

Throughout this paper, the directional parameters related to the EMRI source are fixed as $\theta_s=\pi/3,~\phi_s=\pi/2$, $\theta_l=\pi/4,~\phi_l=\pi/4$,
and the initial orbital separation $r_0$ is adjusted to experience one-year adiabatic evolution before the final plunge $r_{\text{end}}=r_{\text{ISCO}}+0.1~M$.
We consider the EMRI system with $m_p=10~M_{\odot}$, $M=10^6~M_{\odot}$, the different Proca masses $\mu$, and the luminosity distance $d_L$ can be changed freely to vary the SNR of the signal.
\subsection{Fluxes of the Proca field}\label{fluxes}
In this subsection, we study the behaviors of the Proca fluxes radiated by the EMRI system.
In our calculation of the fluxes, it was found that the fluxes of the Proca field are equal to zero at the infinite when the orbital frequencies are less than the mass of the Proca field $\omega_m<\mu$. This phenomenon is similar to the case of massive scalar fields \cite{Berti:2012bp,Barsanti:2022vvl}.
The fluxes at the horizon are always present for the arbitrary orbital frequency,
which play a role in the orbital evolution of EMRI  during the entirety of the inspiral phase.

The total Proca fluxes can be computed by summing over the modes $(l,m)$, where the modes $(l,m)$ vary in the ranges of  $1\leq l\leq l_{\rm max}$ and $1\leq m \leq l$, and the highest mode is set as $l_{\rm max}=10$.
For the vector emission, the relative differences of the fluxes for different vector filed cases are reported in Table \ref{Fluxvalues}.
For the fluxes at the infinity, one can find that the relative difference of the fluxes for the modes $(l_{\rm max}, l_{\rm max}+1)$ is less than $10^{-2}\ \%$ when the mass of Proca field is in the range of $0\leq \mu \leq 0.05$.
However, for the fluxes near the horizon, the relative difference of the fluxes for the succession modes is always less than $10^{-2}\ \%$ when the mass of the Proca field is $0\leq \mu \leq 0.5$.
For larger values of the Proca mass $\mu=0.5$, the relative ratio of the Proca flux at infinity  is $\leq93\%$, so the contribution of the higher mode $l=11$ on Proca flux is not neglected when the secondary is closing to the MBH.

\begin{table*}[htbp!]
\centering
\begin{tabular}{cccccccc}
\hline
$ \mu $ &   $r_0/M$ &  $\dot{E}^{l_{\text{max}}=10}_\infty$ &$\dot{E}^{l_{\text{max}}=11}_\infty$ & Rel. Diff. &  $\dot{E}^{l_{\text{max}}=10}_H$ &$\dot{E}^{l_{\text{max}}=11}_H$ & Rel. Diff. \\
  \hline
0 &6.5  &$3.598\times10^{-4}$ &  $3.598\times10^{-4}$ & $<10^{-2} \ \%$  &$8.105\times10^{-6}$ &  $8.105\times10^{-6}$ & $<10^{-2} \ \%$ \\
    &12 &$3.031\times10^{-5}$ &$3.031\times10^{-5}$  & $<10^{-2} \ \%$ &$8.526\times10^{-8}$ &$8.526\times10^{-8}$  & $<10^{-2} \ \%$\\
\hline
0.01 &6.5  &$3.600\times10^{-4}$ &$3.600\times10^{-4}$  & $<10^{-2} \ \%$  &$8.067\times10^{-6}$ &$8.067\times10^{-6}$  & $<10^{-2} \ \%$\\
    &12 &  $3.006\times10^{-5}$  &$3.006\times10^{-5}$ & $<10^{-2} \ \%$ &  $8.395\times10^{-8}$  &$8.395\times10^{-8}$ & $<10^{-2} \ \%$ \\
    \hline
0.05 &6.5  &$3.113\times10^{-4}$ &$3.114\times10^{-4}$  & $<10^{-2} \ \%$  &$7.180\times10^{-6}$ &$7.180\times10^{-6}$  & $<10^{-2} \ \%$\\
    &12 &  $2.872\times10^{-7}$  &$2.872\times10^{-7}$ & $<10^{-2} \ \%$ &  $6.612\times10^{-8}$  &$6.612\times10^{-8}$ & $<10^{-2} \ \%$ \\
    \hline
0.5 &6.5  &$3.022\times10^{-14}$   & $4.487\times10^{-13}$& $<93 \ \%$   &$9.669\times10^{-8}$   & $9.669\times10^{-8}$& $<10^{-2} \ \%$ \\
    &12   &$0$   &$0$ &  $<10^{-2} \ \%$  &$6.187\times10^{-12}$   &$6.187\times10^{-12}$ &  $<10^{-2} \ \%$\\
\hline
\hline
\end{tabular}
\caption{The vector fluxes (in units of $m_p^2/M^2$)  emitted from the EMRI system for the orbital radius $r_0=6.5M$ and $r_0=12M$  are listed, which consider the different vector fields with the masses $\mu \in \{0,0.01,0.05,0.5\}$.
The fifth column and eighth column are the relative difference of the vector flux for $l_{\text{max}}=10$ and $l_{\text{max}} = 11$, which correspond to the cases at the horizon and at the infinity.
Note that the parameters $\mu=0$ and $\mu\neq0$ denote the cases of the massless vector field and the Proca field. }\label{Fluxvalues}
\end{table*}

\begin{figure}
   \centering
   \includegraphics[width=1.026\columnwidth]{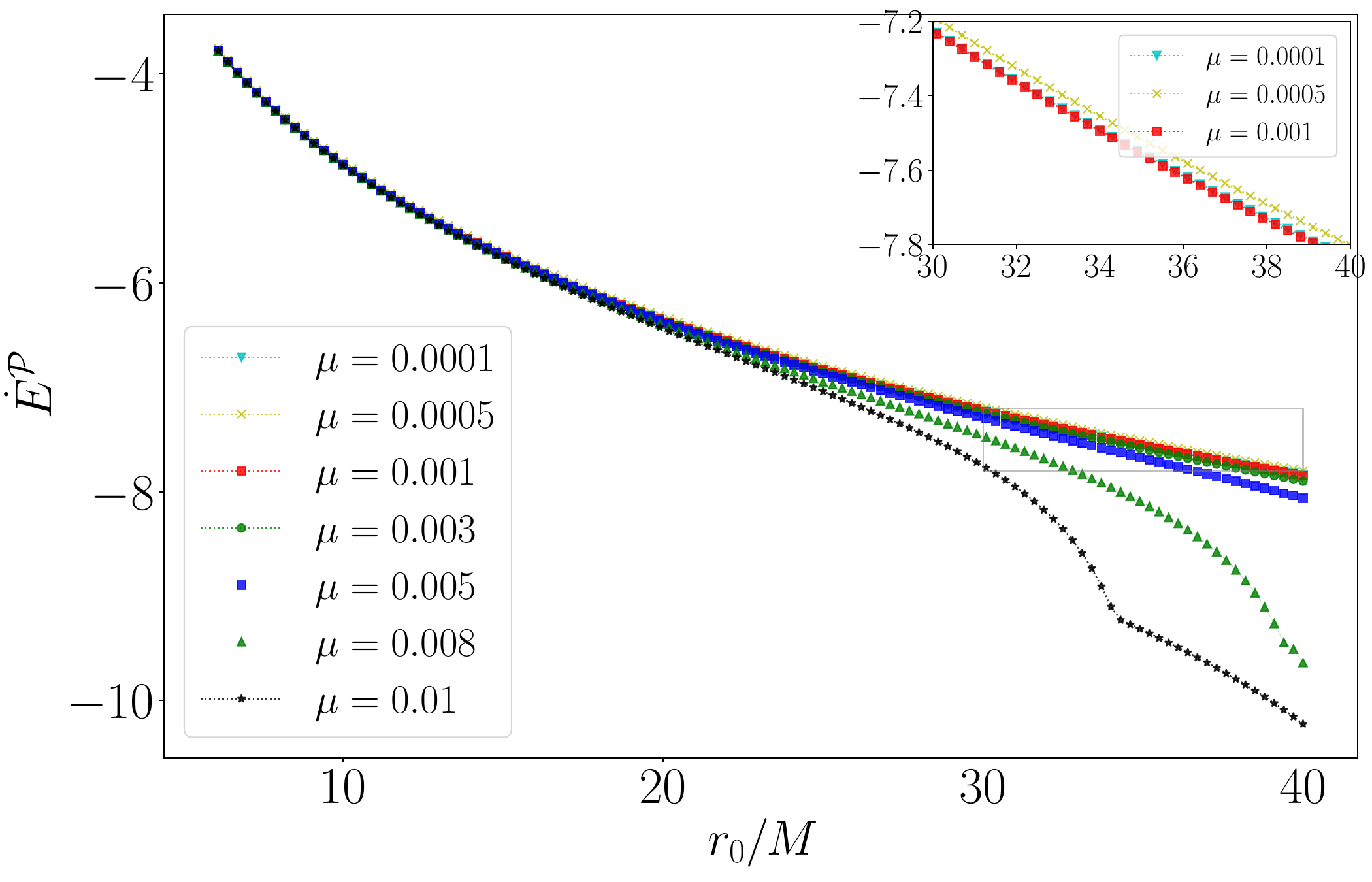}
   \includegraphics[width=1.026\columnwidth]{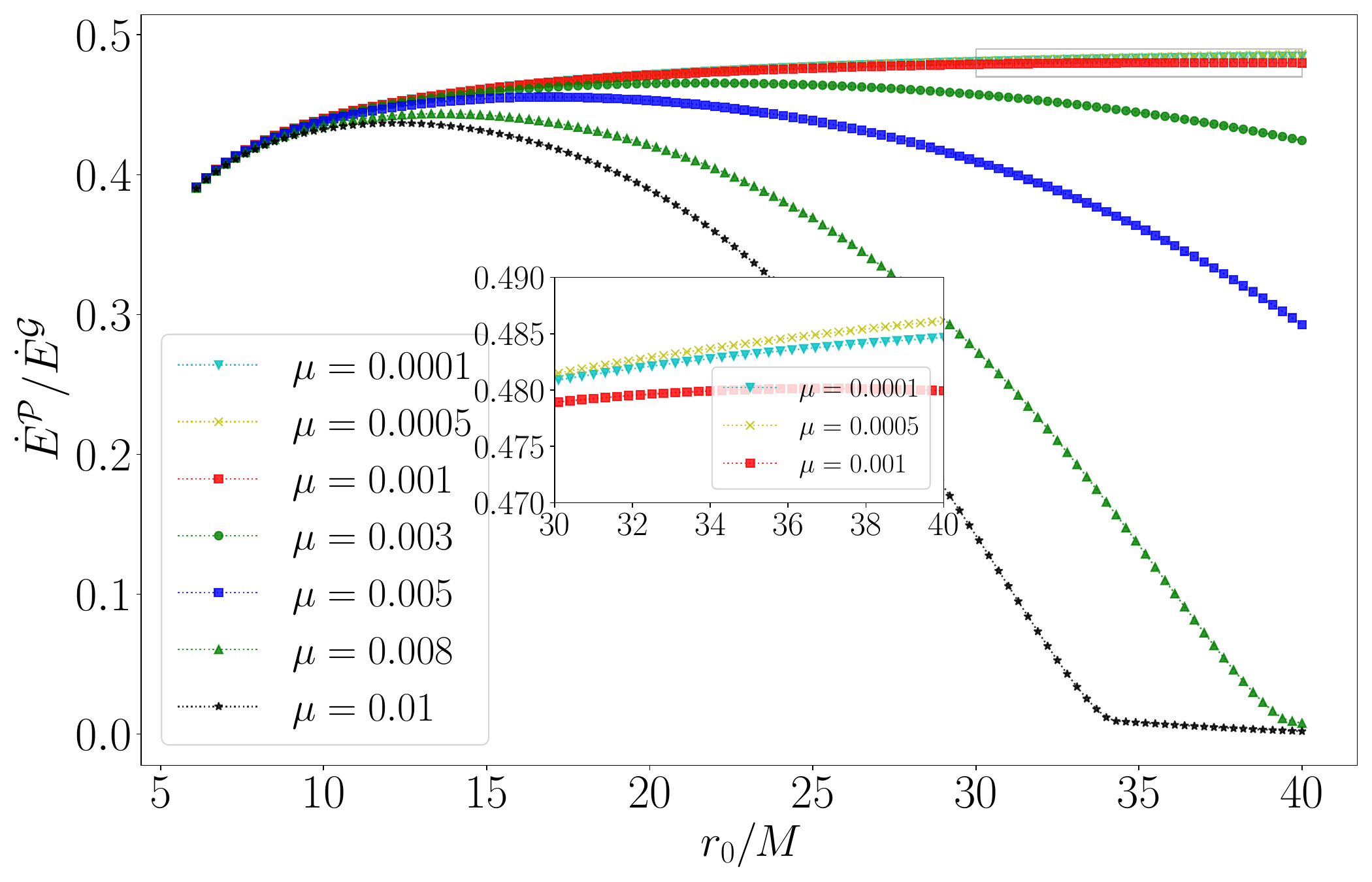}
   \caption{The Proca flux and the ratio between the GW fluxes in the GR and in the Einstein-Proca gravity as a function of the orbital radius $r_0/M$ and the Proca mass $\mu$ are plotted,
    which consider the cases of different Proca fields with a mass $\mu \in \{0.0001,0.0005,0.001,0.003,0.005,0.008,0.01\}$.
    The subfigures are the zoom magnified pictures in the ranges $r_0/M\in[30,40] $ of orbital radius.
    Note that these fluxes have the units of $m_p^2/M^2$.
    }\label{energyProca}
\end{figure}

Fig. \ref{energyProca} displays the Proca and gravitational fluxes ratio $\dot{E}^\mathcal{P}/\dot{E}^\mathcal{G}$ and the logarithm $\log_{10}(\dot{E}^\mathcal{P})$  of the Proca flux as a function of orbital radius $r_0/M$, considering the cases of different Proca masses, and the multipole contribution of the fluxes includes the order $l_{\rm max}=10$.
One can find that the Proca flux is increasing as the binary objects shrink slowly up to near the ISCO, and the quantitative magnitudes of the Proca fluxes are almost the same with the massless case when the Proca mass is  $\mu< 0.003$.
From the bottom of Fig. \ref{energyProca}. the Proca flux near the ISCO can reach $\sim40\%$ of the gravitational flux, however, the Proca fluxes decreases when the Proca mass is increasing,
which would cause a sharp decline for the more massive Proca field accorrding to two subfigures in Fig. \ref{energyProca}.
Therefore, one can forecast that the deviation of the EMRI dynamics evolution between the Einstein-Proca case and the standard GR case would be obvious when the vector field becomes heavier, however, the effect of Proca field on the EMRI waveform would be weaken a little due to that the Proca flux is supressed by its mass.

\begin{figure}
    \centering
    \includegraphics[width=1.026\columnwidth]{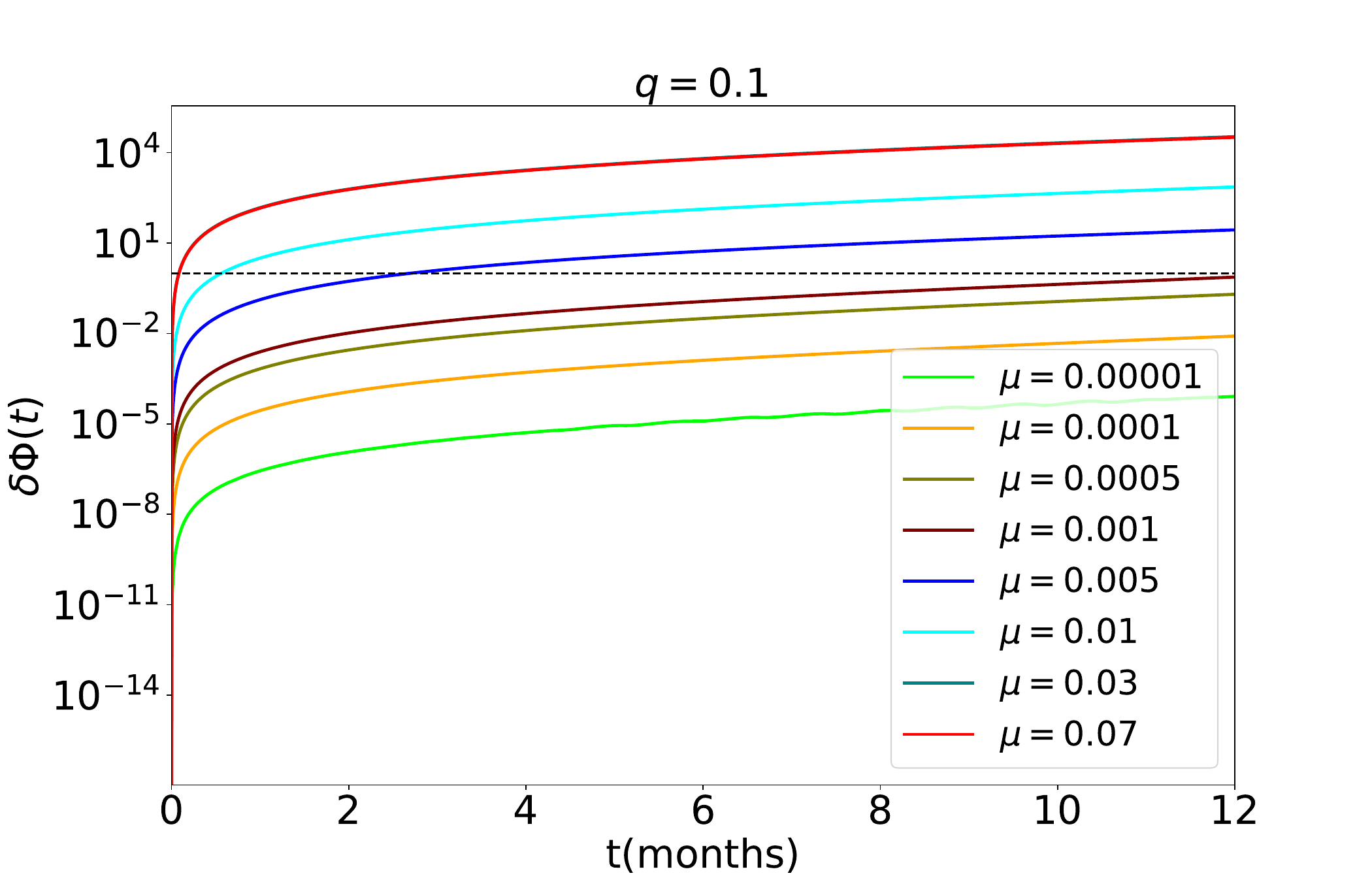}
    \includegraphics[width=1.026\columnwidth]{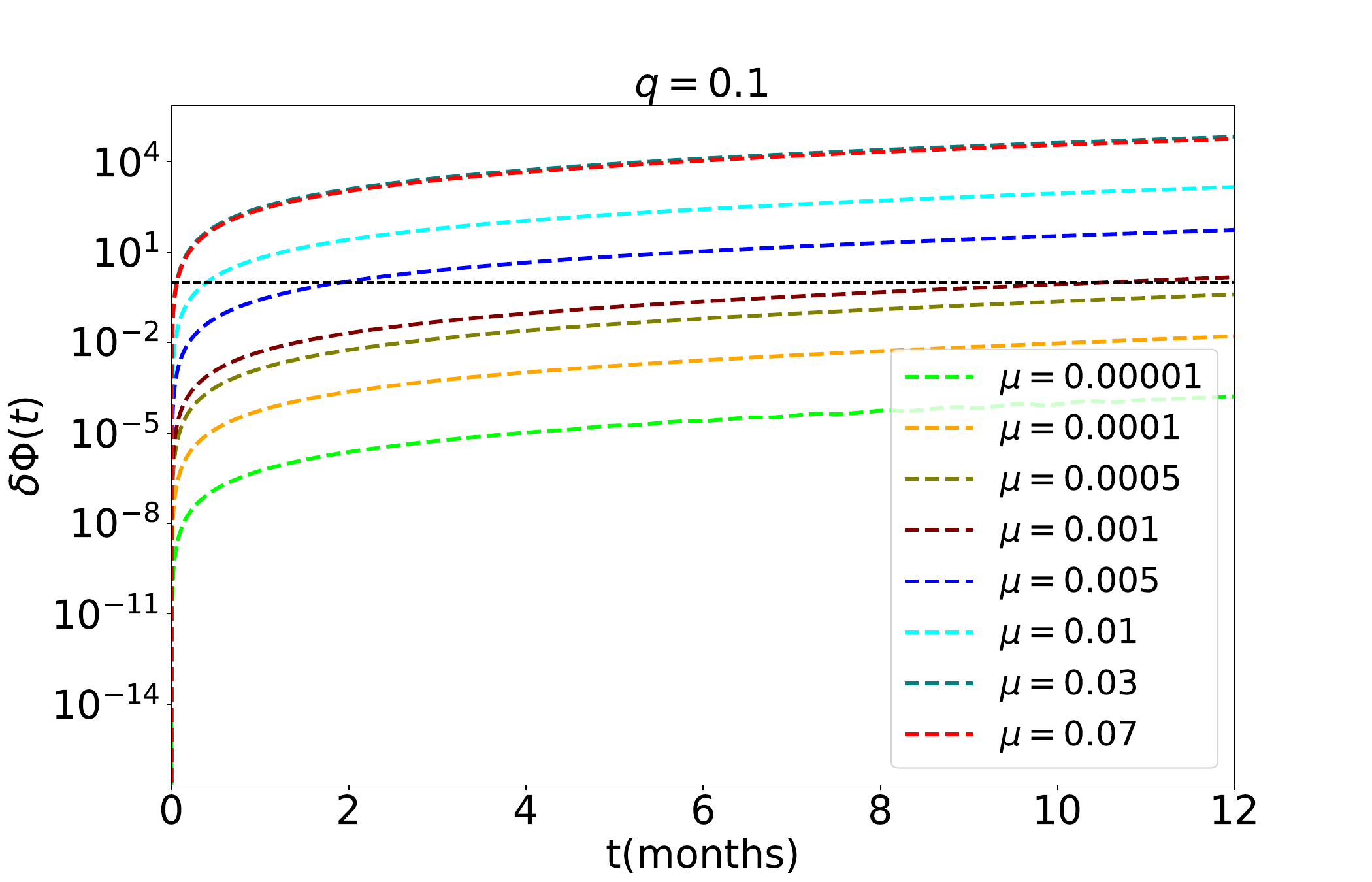}
    \caption{The differences of EMRI orbital phases, for the massive and massless vector fields with the vector charge $q=0.1$ (top panel) and for the massive vector and standard GR cases (bottom panel), as a function of the observation time are plotted. The top panel contains some cases of the Proca fields with the masses $\mu \in \{0.00001,0.0001,0.0005,0.001,0.005,0.01,0.03,0.07\}$,
     the bottom panel considers the comparison between the  standard case and  different charged massive vector cases with  $q\in \{0.00001,0.0001,0.0005,0.001,0.005,0.01,0.03,0.07\} $.
    The horizontal black lines denote to the threshold value of dephasing for the top and bottom panel.  }\label{fig2:dephasing}
\end{figure}

\begin{figure}
    \centering
     \includegraphics[width=1.026\columnwidth]{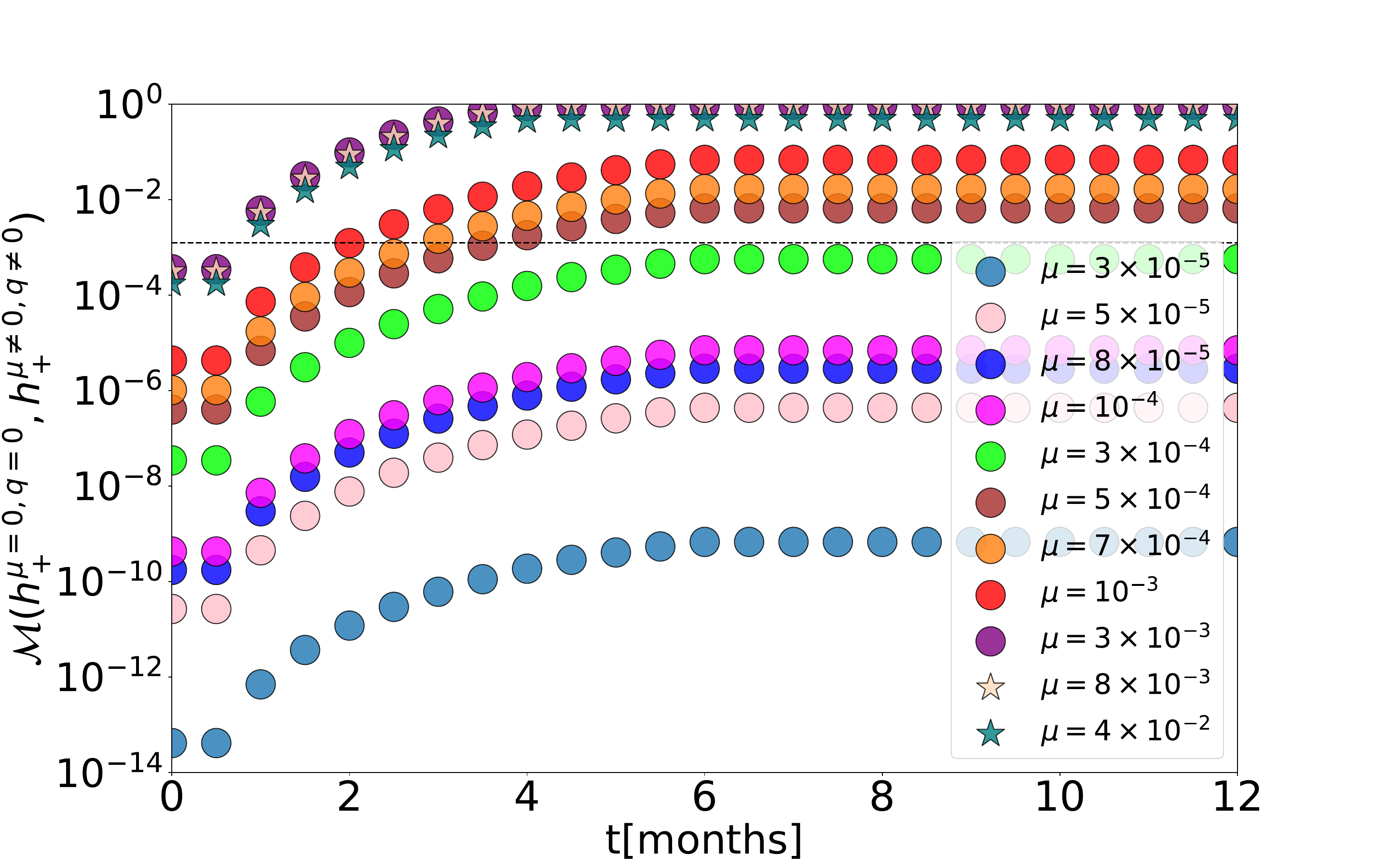}
     \includegraphics[width=1.026\columnwidth]{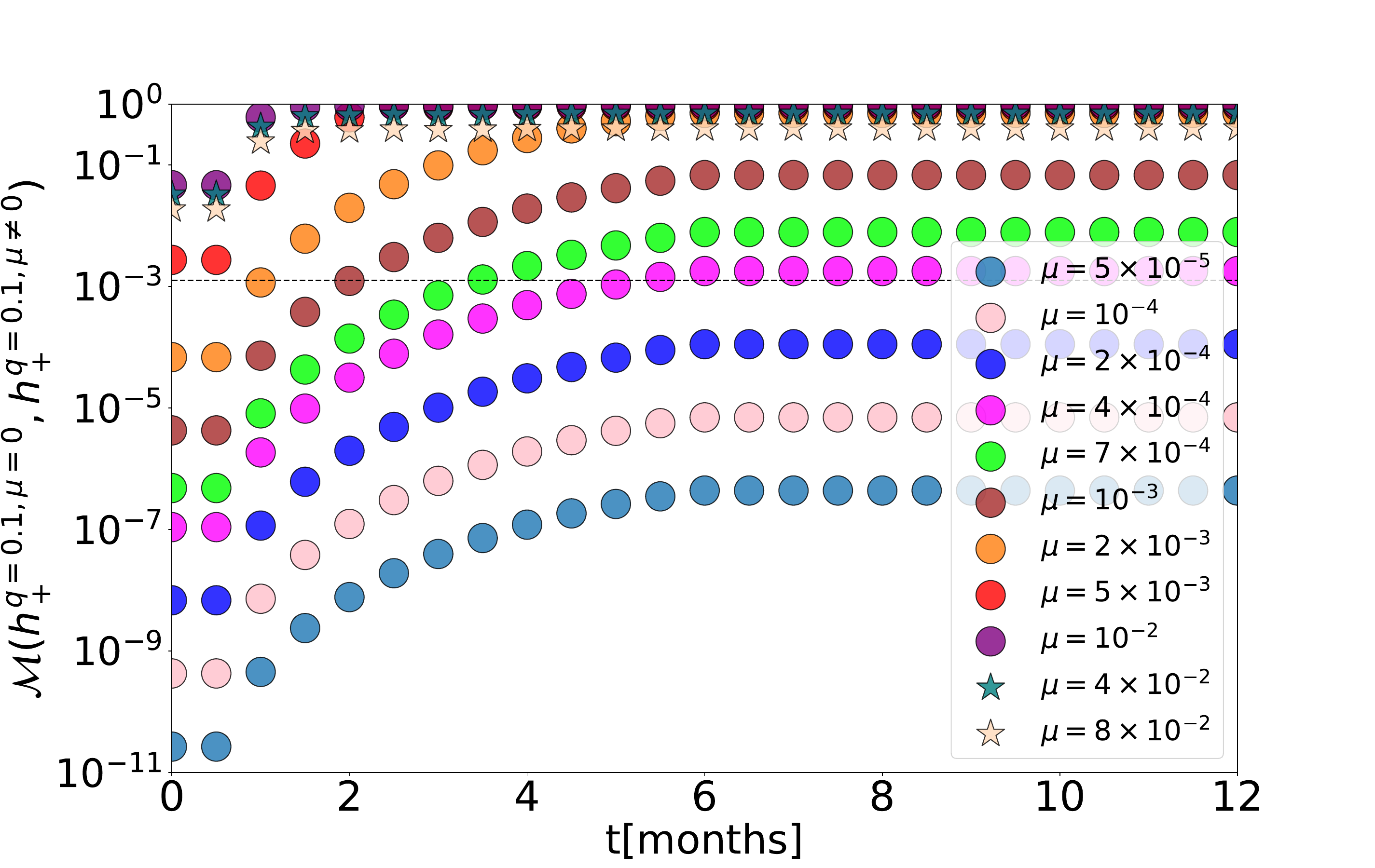}
     \caption{The mismatches between the massless vector field or standard GR cases and the Proca field cases  are plotted, where the  parameters of the  vector fields carried by the smaller objects are  set as $\left( \{ \mu=0,q=0\}, \{\mu\neq0,q\neq0 \} \right)$ and $\left( \{ \mu=0,q=0.1\}, \{\mu\neq0,q=0.1 \} \right)$, and the other parameters are set as $\{r_0=16, q=0.05, m_p=10 \Msun, M=10^6 \Msun\}$.
    Note that the horizontal black lines denote to the threshold value of mismatch for the top and bottom panels.
}\label{fig3:mismatch}
\end{figure}

\subsection{dephasing and mismatch}\label{dephasing}
To evaluate the effect of the Proca field on the EMRI dynamics, we analyze the differences in the orbital phases between the massive and massless vector cases.
We compute the dephasing \eqref{dephasing:formula} as a function of the observation time in the Fig. \ref{fig2:dephasing}, where the dashed horizontal line denotes the threshold value $\delta \Phi_{\rm min}=1$ radian discerned by LISA.
Note that the difference of two waveform phases is generated by the EMRI systems including the secondary objects with two vector fields $(q=0.1,\mu=0)$ and $(q= 0.1,\mu\neq0)$, where the length of waveforms are both set as one year.
According to the top panel of Fig. \ref{fig2:dephasing}, the minimum value of the Proca mass for the charge $q=0.1$ case distinguished by LISA is $\mu_{\rm min}\sim 0.0005$, the dephasing becomes bigger when the Proca field is heavier.
Additionally, comparing with the case of $\mu=0.03$, we find that the mismatch would decrease slightly for the case of $\mu=0.07$.
The bottom panel of Fig. \ref{fig2:dephasing} displays the dephasing beteween the Proca field cases with charge $q=0.1$ and the standard GR case,
the dephasing would be slightly bigger comparing to the dephasing between the Proca field case and the massless vector field case.
The analysis reveals that the lighter Proca field with the mass $\mu\gtrsim 0.0005$ could result in the
observable modification effect in the EMRI phase evolution, but the rigorous constraint on the Proca field effect needs to consider the correlation of waveform parameters, such as the mismatch and Fisher matrix analysis in the following section.

In order to assess quantifiably the effect of the Proca mass on the EMRI waveforms, the mismatch as a function of the observation time is plotted in Fig. \ref{fig3:mismatch},  which takes into account the differences of two waveforms between the standard GR case or the massless field case and the Proca field case.
It should be noted that the horizontal dashed black lines correspond to the threshold value distinguished by LISA in Fig. \ref{fig3:mismatch}.
Overall, the mismatches grow visibly when the Proca mass is bigger for the fixed vector charge $q=0.1$.
The upper panel of the Fig. \ref{fig3:mismatch} shows the mismatches for the EMRI systems including the secondaries with the vector field parameters $(q=0,\mu=0)$ and $(q=0.1, \mu\neq0)$,
one can see that the minimum value of the Proca mass recognized by LISA is  $\mu_{\rm min}\sim 5\times 10^{-4}$  for the vector charge $q=0.1$ case.
From the bottom panel of the Fig. \ref{fig3:mismatch}, the mismatches between the massless vector field case and the Proca field case are plotted, which consider the smaller objects with $(q=0.1,\mu=0)$ and $(q=0.1, \mu\neq0$. As shown in the bottom panel of the figure, LISA can distinguish the effect of the Proca field with the mass as small as $\mu_{\rm min}\sim 4\times 10^{-4}$.

Our results indicate that, for the Proca field with the mass $\mu\sim 4\times 10^{-4}$
and the charge $q=0.1$, LISA can distinguish the EMRI waveforms including the modification of the massive vector field from the waveforms within the massless vector filed case and the standard GR case. Additionally, one can see that the mismatch would begin to decrease when the Proca mass $\mu$ exceeds a certain value according to two panels.
We observe that for smaller Proca masses, including the vector field mass into the EMRI system introduces additional degrees of freedom for the vector field case (three propagating degrees), leading to an increase in the Proca flux compared to the massless vector field scenario (two propagating degrees). However, as the Proca mass becomes excessively large and surpasses a certain threshold, the Proca flux is suppressed by the mass. Specifically, as shown in Fig. \ref{fig2:dephasing} and Fig. \ref{fig3:mismatch}, this results in a slight reduction in the dephasing and mismatch.
These analyses can be the criterion
ruling out some parameters with the EMRI observation, the severe constraint on the Proca field needs to consider the parameter estimation using the FIM method in the following subsection.

\begin{figure}
\centering
\includegraphics[width=1.1\columnwidth]{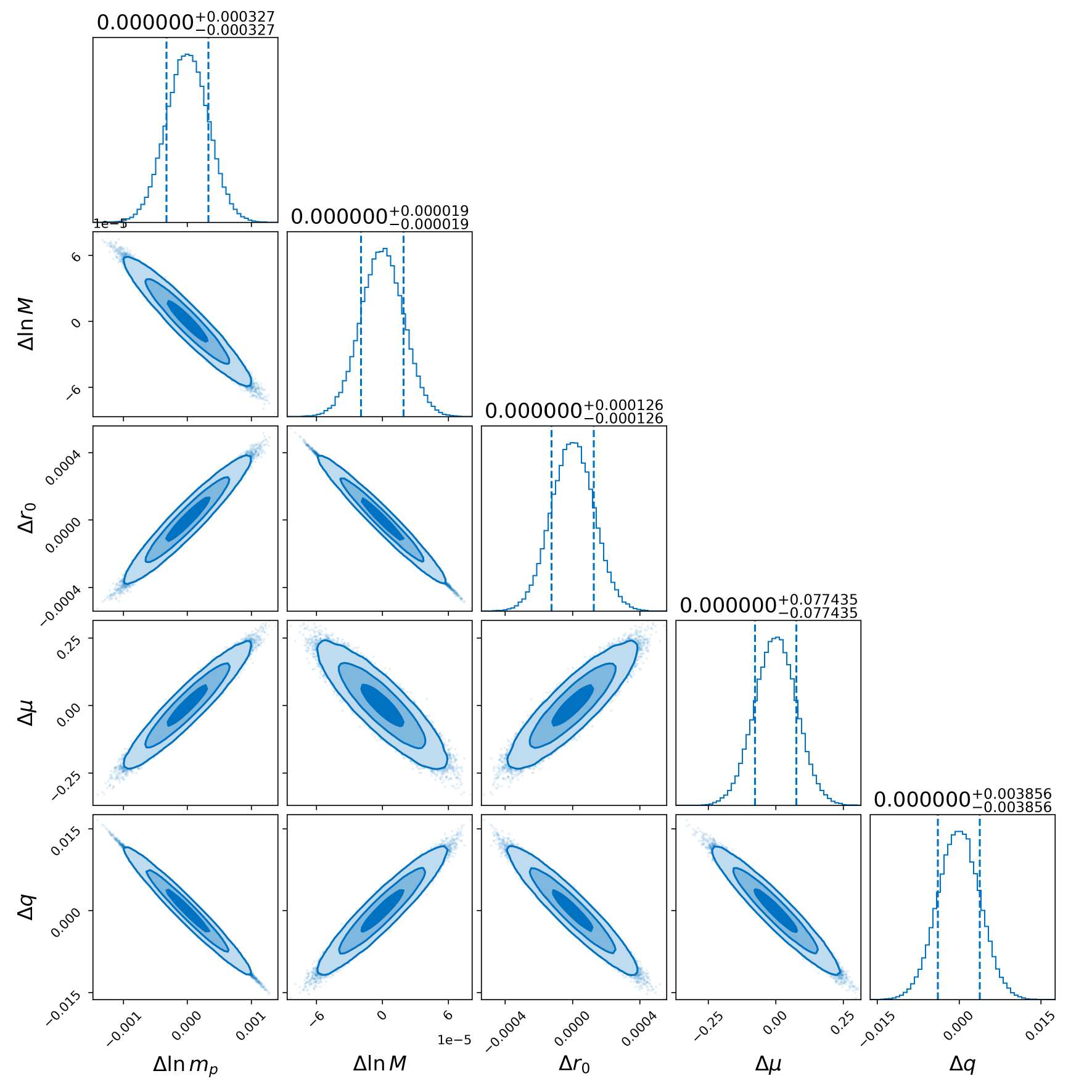}
  \caption{The corner plot for the probability distribution for the Proca field mass, the orbital radius, vector charge, secondary body mass, and primary body mass, $(\mu=0.01,r_0=16,q=0.05, m_p=10 \Msun,M=10^6 \Msun)$, which is inferred from one-year observation of EMRI. Vertical lines show the $1\sigma$ interval for each source parameter. The contours correspond to $68\%$, $95\%$ and $99\%$ probability confidence intervals, and
  the symbol $``\Delta"$ means the difference between the distribution for the parameter value and the true parameter value.
}\label{fig4:corner}
\end{figure}

\begin{figure}
    \centering
    \includegraphics[width=1.1\columnwidth]{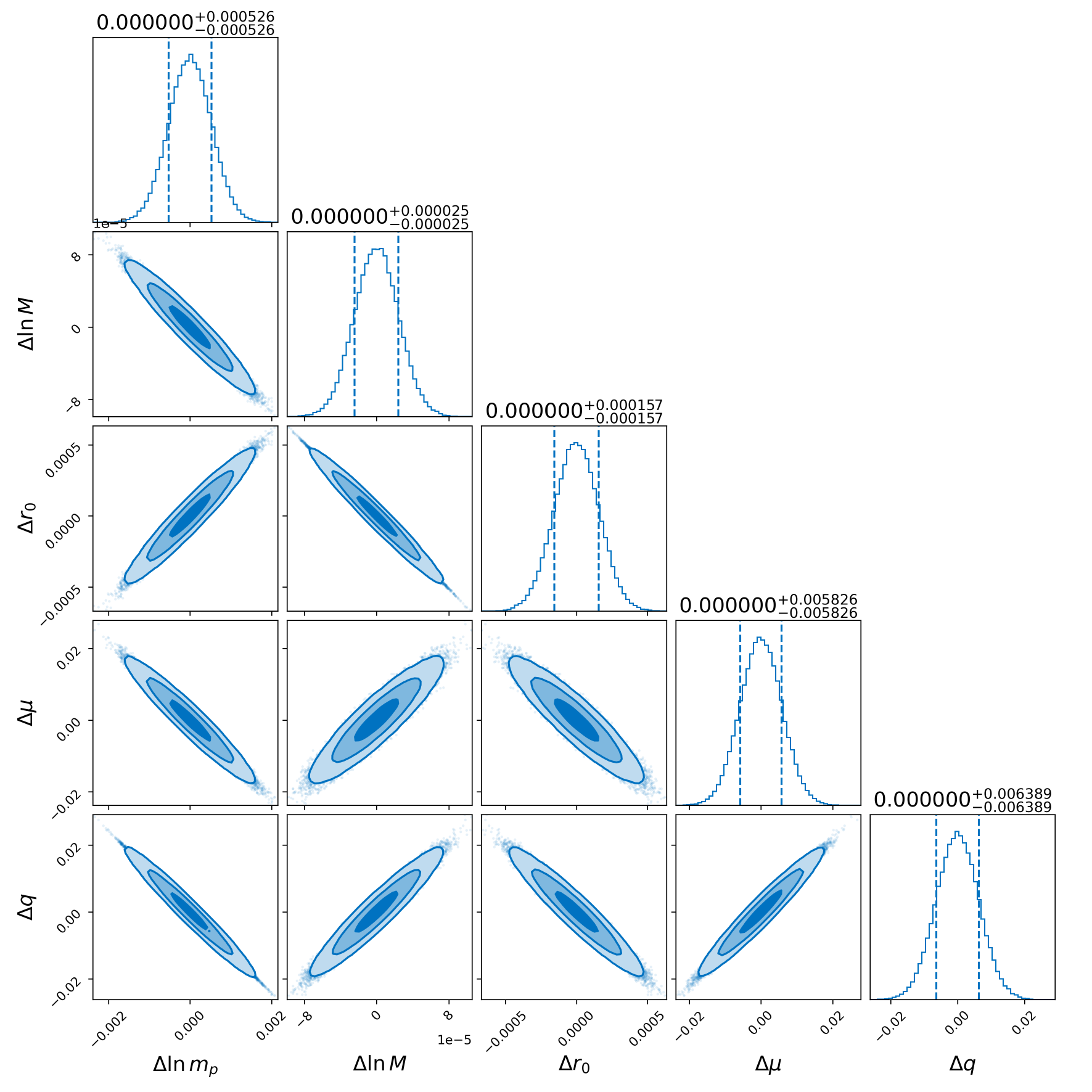}
    \caption{The corner plot for the probability distribution for the Proca field mass, the orbital radius, vector charge, secondary body mass, and primary body mass, $(\mu= 0.02, r_0=16, q=0.05, m_p=10 \Msun,M=10^6 \Msun)$, which is inferred from one-year observation of EMRI. Vertical lines show the $1\sigma$  interval for each source parameter. The contours correspond to $68\%$, $95\%$, and $99\%$ probability confidence intervals,
    and the symbol $``\Delta"$ means the difference between the distribution for the parameter value and the true parameter value.} \label{fig5:corner}
\end{figure}

\subsection{Constraint on the Proca field}\label{fim}
In this subsection, we plan to adopt the FIM method to compute the constraint on the mass of the Proca field using the EMRI waveform modified by the massive vector field.

The measuring errors of EMRI parameters are listed in the Table \ref{tab:fim} for the cases of the Proca
field masses $\mu=0.01~(\sim 8.39\times 10^{-19} \rm eV)$ and $\mu=0.02 ~(\sim 1.68\times 10^{-18} \rm eV)$, in which the vector charges of secondaries are set as $q=0.05$.
As shown in the table, the constraint accuracy of the Proca field mass can reach a fraction of
$\sigma_\mu \sim10^{-2} ~(\sim 8.39\times 10^{-19} \rm eV)$ for the case of $\mu=0.02$, which would slightly be improved comparing with the case of $\mu=0.01$. The effect of the massive vector field on the EMRI waveform  for the Proca mass $\mu=0.02$ case becomes more obvious though the fluxes are slightly smaller compared with the Proca mass $\mu=0.01$ case, the more massive vector field case is more deviation from the GR case.
Therefore, the Proca mass $\mu=0.02$ case can be constrained more rigorously.
Additionally, the constraint on the vector charge can reach $\sigma_q \sim 3.86\times 10^{-3}$ when including the massive vector field in the EMRI system, the measurement error is $\sim8.59\times 10^{-2}$ if the EMRI system exists the massless vector radiation \cite{Zhang:2023vok}.
So the bound of vector charge can become more rigorous when considering the massive vector field for the EMRI signal.
In the end, since the analysis with EMRI waveforms is very delicate, so we also assess the stability of FIM using the method \cite{Speri:2021psr,Maselli:2021men,Zi:2022hcc},
the detailed procedure can refer to the appendix \ref{APPB:stability}.
Finally, in Fig. \ref{fig4:corner}, we display the probability distribution of the Proca field mass, the orbital radius, vector charge, secondary body mass, and primary body mass, where the mass and charge of the Proca field are set as $\mu=0.01$ and charge $q=0.05$.
The constraint result for the same parameters for the Proca mass $\mu=0.02$ and charge $q=0.05$ case is also presented in Fig. \ref{fig5:corner}.
As shown in Fig. \ref{fig4:corner} and Fig. \ref{fig5:corner}, the correlation between the Proca mass with other parameters for the case $\mu=0.01$ is inverse to the case $\mu=0.02$.

\begin{table*}
\centering
\begin{tabular}{c|c|c|c|c|c}
\hline
 $ \mu$ & $\sigma_{\mu}/\mu$  &$\sigma_q$    &$\sigma_{m_p}/m_p$  &$\sigma_{M}/M$ & $\sigma_{r_0}$ \\
\hline
$0.01$ &$774 \%$ & $3.86\times10^{-3}$ & $3.27\times10^{-4}$    &$1.90\times10^{-5}$ &$1.26\times 10^{-4}$ \\
\hline
$0.02$  & $29.6\%$ & $6.39\times10^{-3}$   &$5.26\times10^{-4}$   &$2.50\times10^{-5}$   &$1.57\times 10^{-4}$ \\
\hline
\end{tabular}
\caption{Measurement errors of the source parameters using the EMRI waveform modified by the Proca field are listed, the parameters are set as $m_p=10~M_{\odot}$, $M=10^6~M_{\odot}$, $q=0.05$ and the SNR of the signal is set as $150$.  }\label{tab:fim}
\end{table*}

\section{Conclusion}\label{sumup}
In this paper, we consider the vector charge and mass carried by the secondary object in the EMRI,
where the waveform from the inspiral evolution including the extra dipole radiation can be regarded as the probe of detecting the ultralight vector field.
We first solve the homogeneous Proca equation and the Regge-Wheeler equation with the
ingoing and outgoing boundary conditions, and derive the source term of a point particle on the circular orbits for the Proca field, then derive the energy flux formulas of the Proca field at infinity and near the horizon.
Using the independent homogeneous solutions and the source term, one can compute the energy fluxes for the Proca emission and the gravitational emission.
We observe that the Proca flux exhibits a non-monotonic behavior with respect to the mass of the vector field. Specifically, the flux increases with the Proca mass in the regime of small vector field masses, whereas it decreases as the Proca mass continues to grow beyond a certain threshold.

Under the framework of adiabatic approximation, we evolve the circular orbital parameters of a point particle on the equatorial plane, then compute the EMRI waveform using the quadrupole approximation formula.
Using the adiabatic EMRI waveforms, we can analyze the effect of the massive vector field on the
EMRI waveforms by computing the dephasing and mismatch.
Our results show that LISA can distinguish the EMRI waveforms for the Proca field with mass $\mu\sim 4\times 10^{-4}$ from the massless vector field cases.
Finally, we carry out the parameter estimation of the EMRI source including the modification by the Proca field.
According to the constraint computed by the FIM method, the measurement error of the Proca field mass can be controlled within the fraction of $\sim10^{-2}~(\sim 8.39\times 10^{-19} \rm eV)$ using the EMRI observation. The constraint on the Proca mass is anti-correlated with the masses $M, m_p$ of MBH and the secondary body, and the correlation would be interesting to include the spin of MBH and the eccentric orbit in the future.

In this paper, the constraint on Proca mass using the FIM method would not be reliable for future EMRI data analysis. The full Bayesian analysis based on Monte Carlo Markov Chain simulations should be performed to constrain the Proca mass more rigorously, but the method is very exprensive to compute a mass of EMRI waveforms, so the bounded result obtained with Fast EMRI Waveforms \cite{Katz:2021yft} may be feasible in the future. Our current model only focuses on the circular orbits on the equatorial plane, because the EMRI orbits are eccentric and inclined, and the FIM and Bayesian analyses both reply on the accurate waveform model, so the full relativistic EMRI template should be considered the effect of Proca field on generic orbits in the adiabatic approximation.
Additionally, the statistical analysis requires that the EMRI template can generate waveform rapidly, so the exploitation of quadrupolar templates also should be appreciated.
Furthermore, the inclusion of MBH spin would trigger the superradiance phenomenon, which also modifies the EMRI fluxes and orbital evolution \cite{Fell:2023mtf}. This approach for constraining the Proca field using EMRIs around rotating MBH warrants further investigation.

\begin{acknowledgments}
This work was supported by the China Postdoctoral Science Foundation under Grant No. 2023M742297 and No. 2023M731137. T. Z. is also funded by the National Natural Science Foundation of China with Grants No. 12347140 and No.12405059.

\end{acknowledgments}

\begin{figure}
    \centering
     \includegraphics[width=1.026\columnwidth]{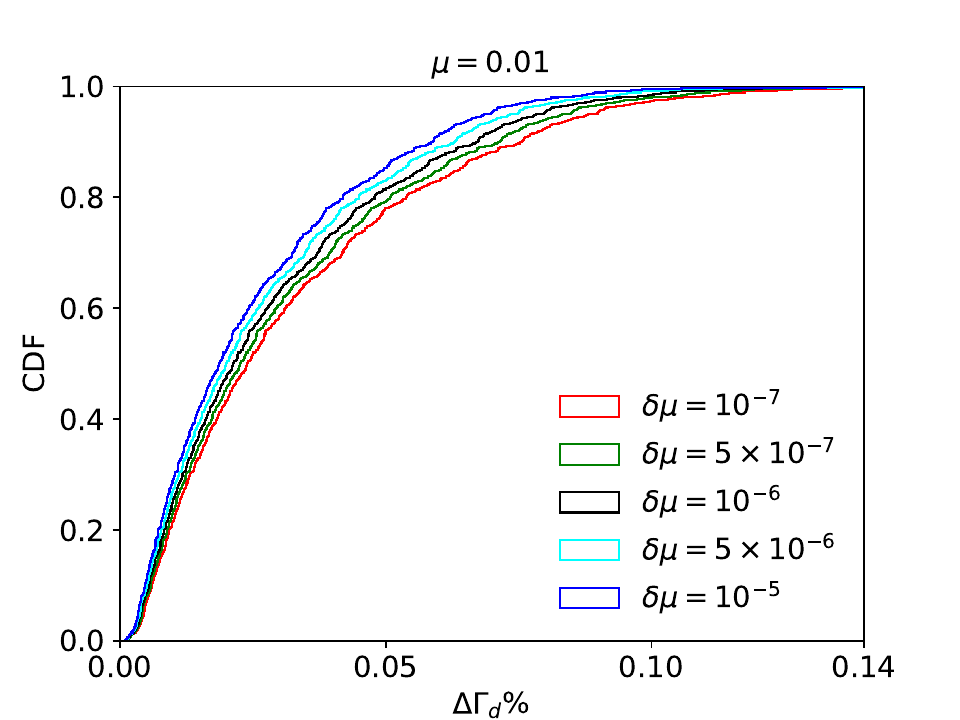}
     \includegraphics[width=1.026\columnwidth]{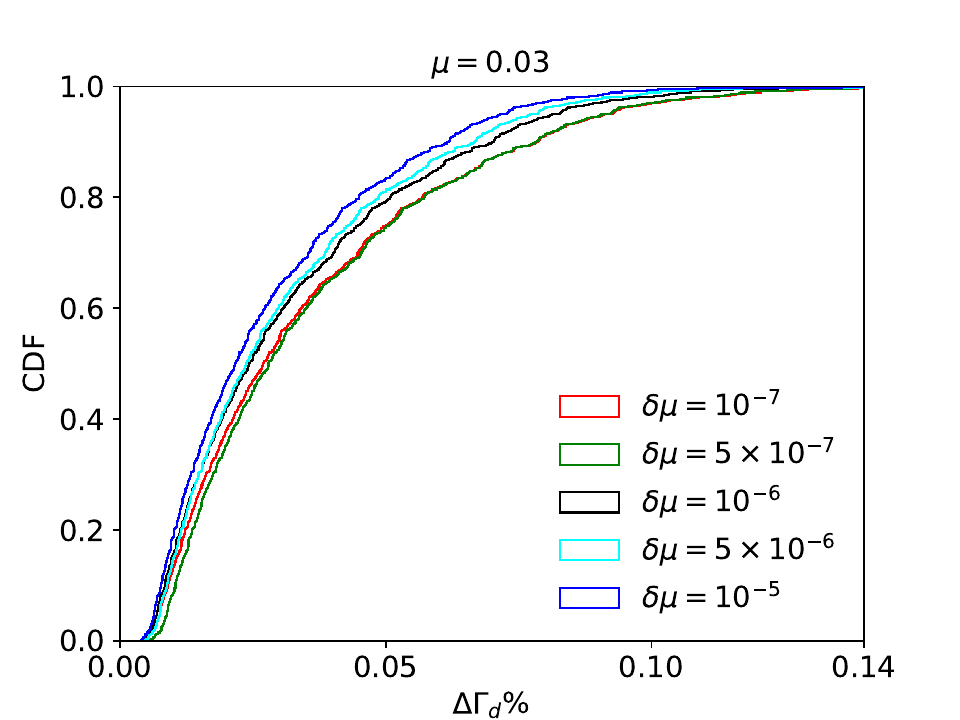}
     \caption{The cumulative distribution of the maximum relative error for FIM and perturbed FIM with the element in a uniform distribution.
     Different color cures denote to the numerical derivative spacing for the calculation of FIM, the elements $U$ of deviation matrix are the uniform distribution, $U\in [-10^{-6},10^{-6}]$.
}\label{fig6:CDF}
\end{figure}

\appendix
\section{Energy fluxes formulas for the massive vector field}\label{appendix1}
From the lagrangian \eqref{Lagrangian},
one can compute an effective canonical stress-energy tensor $T^C_{\mu\nu}$ for the vector perturbations using the Noether theorem, which can be reduced to the following expression
\begin{equation}
4\pi T^C_{\mu\nu}= \frac{\partial \mathcal{L}}{ \partial A^{\alpha,\mu} }A^{\alpha}_{~~,\nu}-g_{\mu\nu}\mathcal{L} ,
\end{equation}
where $A^{\alpha,\mu}\equiv g^{\mu\nu}\frac{\partial A^\alpha}{\partial x^\nu}$ and $A^\alpha_{~~,\nu}\equiv \frac{\partial A^\alpha}{\partial x^\nu}$.
Using the field equations to further simplify our calculations, we find
\begin{equation}
4\pi T^C_{\mu\nu}= F_{\mu\lambda}\partial_\nu A^\lambda-g_{\mu\nu}\mathcal{L},
\end{equation}
there is a general procedure for constructing a symmetric stress tensor
\begin{eqnarray}
4\pi T_{\mu\nu}&=& F_{\mu\lambda}\partial_\nu A^\lambda-F_{\mu}^{~~\lambda}\partial_\lambda A_\nu-A_\nu\partial_{\lambda}F_{\mu}^{~~\lambda}-g_{\mu\nu}\mathcal{L} \nonumber\\& +&\frac{\partial}{\partial x^{\lambda}}(F_{\mu}^{~~\lambda} A_\nu) \nonumber\\
&=&g^{\alpha\beta}F_{\mu\alpha}F_{\nu\beta}+\mu^2 A_{\mu}A_{\nu}-g_{\mu\nu}\mathcal{L}.
\end{eqnarray}
The vector fluxes are then given by integrating the components $T_{ti}$, which provide the energy that flows across the unit surface orthogonal to the axis $x^i$ per unit time \cite{Martel:2003jj,Zhu:2018tzi},
\begin{eqnarray}\label{dedt}
\left\langle\frac{d E}{dt}\right\rangle_{H,\infty}&=&\lim_{r\to r_h,\infty}\int T_{ti}n^i r^2 d\Omega=
 \nonumber \\ &-& \lim_{r\to r_h,\infty}\epsilon\int T_{tr}r^2fd\Omega,
\end{eqnarray}
where $\epsilon$ is $1$ corresponding to the flux at the infinity and  $\epsilon$ is  $-1$  corresponding to the flux near the horizon.
\subsection{energy flux at infinity}
The vector potential at infinity can be expressed by
\begin{eqnarray}\label{inf1}
A_t&=\frac{\mathcal{A}^{1+}}{r} e^{-i\omega_m t}e^{i \sqrt{\omega_m^2-\mu^2} r_\ast}r^{\frac{i\mu^2}{\sqrt{\omega_m^2-\mu^2}}}Y^{lm},  \nonumber \\
A_r&=\frac{\mathcal{A}^{2+}}{rf} e^{-i\omega_m t}e^{i \sqrt{\omega_m^2-\mu^2} r_\ast}r^{\frac{i\mu^2}{\sqrt{\omega_m^2-\mu^2}}}Y^{lm}, \nonumber \\
A_\theta&=\frac{1}{l(l+1)}\left[\mathcal{A}^{3+}Y^{lm}_{,\theta}
+\frac{1}{\sin\theta}\mathcal{A}^{4+}Y^{lm}_{,\phi} \right]
\nonumber \\ &e^{-i\omega_m t}e^{i \sqrt{\omega_m^2-\mu^2} r_\ast}r^{\frac{i\mu^2}{\sqrt{\omega_m^2-\mu^2}}}, \nonumber\\
A_\phi&=\frac{1}{l(l+1)}\left[\mathcal{A}^{3+}Y^{lm}_{,\phi}-\sin\theta\mathcal{A}^{4+}Y^{lm}_{,\theta} \right]
 \nonumber\\ & e^{-i\omega_m t}e^{i \sqrt{\omega_m^2-\mu^2} r_\ast}r^{\frac{i\mu^2}{\sqrt{\omega_m^2-\mu^2}}}.
\end{eqnarray}
Based on the Lorenz condition, we can obtain the following relationship between the $\mathcal{A}^{1+}$ and $\mathcal{A}^{2+}$
\begin{equation}
\omega_m \mathcal{A}^{1+}+\sqrt{\omega_m^2-\mu_v^2}\mathcal{A}^{2+}=0.\label{inf2}
\end{equation}
After plugging the expressions Eqs. \eqref{inf1} and \eqref{inf2} into the energy flux formula Eq. \eqref{dedt}, we can get the energy flux at infinity
\begin{eqnarray}
\lim_{r\to \infty}\left\langle \frac{d E}{dt}\right\rangle &=&  \sum_{l=1}^\infty \sum_{m=1}^l
\frac{\omega_m\sqrt{\omega_m^2-\mu_v^2}(|\mathcal{A}^{4+}_{lm}|^2+|\mathcal{A}_{lm}^{3+}|^2)}{2\pi l(l+1)}
\nonumber \\ & +&\frac{\mu_v^2}{2\pi}\frac{\sqrt{\omega_m^2-\mu_v^2}}{\omega_m}|\mathcal{A}^{2+}_{lm}|^2.
\end{eqnarray}
\subsection{energy flux near the horizon}
At the horizon, the vector potential can be written as following
\begin{eqnarray}\label{hor1}
A_t&=&\frac{\mathcal{A}^{1-}}{r} e^{-i\omega_m t}e^{-i\omega_m r_\ast}Y^{lm}, \nonumber\\
A_r&=&\frac{\mathcal{A}^{2-}}{rf} e^{-i\omega_m t}e^{-i\omega_m r_\ast}Y^{lm}, \nonumber\\
A_\theta&=&\frac{1}{l(l+1)}\left[\mathcal{A}^{3-}Y^{lm}_{,\theta}+\frac{1}{\sin\theta}\mathcal{A}^{4-}Y^{lm}_{,\phi} \right] e^{-i\omega_m t}e^{-i\omega_m r_\ast}, \nonumber\\
A_\phi&=&\frac{1}{l(l+1)}\left[\mathcal{A}^{3-}Y^{lm}_{,\phi}-\sin\theta\mathcal{A}^{4-}Y^{lm}_{,\theta} \right] e^{-i\omega_m t}e^{-i\omega_m r_\ast}, \nonumber\\
\end{eqnarray}
where $Y^{lm}_{,\theta}\equiv \frac{\partial Y^{lm}}{\partial\theta}$ and
$Y^{lm}_{,\phi}\equiv \frac{\partial Y^{lm}}{\partial\phi}$ .
Based on the Lorenz condition, we can get the following relationship between the $\mathcal{A}^{1-}$ and $\mathcal{A}^{2-}$
\begin{equation}\label{hor2}
\mathcal{A}^{1-}-\mathcal{A}^{2-}=0.
\end{equation}
After plugging the expressions Eqs. \eqref{hor1} and \eqref{hor2} into the energy flux formula Eq. \eqref{dedt}, we can get at the horizon
\begin{equation}
\begin{split}
\lim_{r\to r_h}\left\langle \frac{d E}{dt}\right\rangle=&  \sum_{l=1}^\infty \sum_{m=1}^l \frac{\omega_m^2|\mathcal{A}^{4-}_{lm}|^2}{2\pi l(l+1)}+\frac{\omega_m^2|\mathcal{A}^{3-}_{lm}|^2}{2\pi l(l+1)}\\ &
+\frac{l(l+1)+4\mu_v^2}{8\pi}|\mathcal{A}^{2-}_{lm}|^2
\\ &-\frac{i\omega_m}{4\pi}\left(\mathcal{A}^{2-}_{lm} \mathcal{A}^{3-\ast}_{lm}-\mathcal{A}^{3-}_{lm} \mathcal{A}^{2-\ast}_{lm}\right).
\end{split}
\end{equation}

\section{Numerical stability for Fisher information matrix}\label{APPB:stability}
In this section,  we check the stability of FIM using the previous works \cite{Speri:2021psr,Maselli:2021men,Zi:2022hcc}.
First we define ten dimensions of matrix as $\mathbf{R}$,
where the elements are randomly drawn in a uniform distribution $U\in [-10^{-6},10^{-6}]$.
Then we compute the  inverse of $\mathbf{R} + \mathbf{\Gamma}$, and give the
the maximum relative error between the unperturbed and perturbed FIMs, $\Delta \mathbf{\Gamma}_d = \rm max( ((\mathbf{R} + \mathbf{\Gamma})^{-1} - \mathbf{\Gamma}^{-1})/\mathbf{\Gamma}^{-1}) )$.
We can compute the statistical result of maximum error using the above method iteratively, then the cumulative distribution of $\Delta \mathbf{\Gamma}_d$, for two EMRI systems with Proca mass $\mu=0.01$ and $\mu=0.03$, as
a function of derivative spacing  is plotted in Fig. \ref{fig6:CDF}.
From two panels in the figure, we can observe that more than $85\%$ of all maximum relative errors is $\Delta \mathbf{\Gamma_d}\leq 0.1\%$, so the computation of FIM can be judged as stability.

%

\end{document}